\documentclass[sigconf]{acmart}

\usepackage{booktabs} % For formal tables

\usepackage{algorithm}
\usepackage{algpseudocode}
\usepackage{amsmath} 
\usepackage{amsfonts}
\usepackage{multirow}
\usepackage{subfigure}
\usepackage{hyperref}
\usepackage{tabularx}
\usepackage{enumitem}
\usepackage{geometry}% http://ctan.org/pkg/geometry
\usepackage{lipsum}% http://ctan.org/pkg/lipsum
\usepackage{graphicx}% http://ctan.org/pkg/graphicx

\usepackage[font=small,labelfont=bf,textfont=md]{caption}

\setcopyright{iw3c2w3}
\copyrightyear{2019}
\acmYear{2019}

\acmConference[WWW '19]{Proceedings of the 2019 World Wide Web Conference}{May 13--17, 2019}{San Francisco, CA, USA}
\acmBooktitle{ Proceedings of the 2019 World Wide Web Conference (WWW'19), May 13--17, 2019, San Francisco, CA, USA}
\acmPrice{}
\acmDOI{10.1145/3308558.3313607}
\acmISBN{978-1-4503-6674-8/19/05}

\begin{document}

\title{Jointly Learning Explainable Rules for Recommendation with Knowledge Graph}
% \author{$\text{Weizhi Ma}^{\dagger}$, $\text{Min Zhang}^{\dagger}$\footnotemark[*], $\text{Yue Cao}^{\ddagger}$, $\text{Woojeong Jin}^{\ddagger}$, $\text{Chenyang Wang}^{\dagger}$, 
% $\text{Yiqun Liu}^{\dagger}$, 
% $\text{Shaoping Ma}^{\dagger}$, \text{and} {$\text{Xiang Ren}^{\ddagger}$}\footnotemark[*]}
% \author{123}
\author{$\text{Weizhi Ma}^{\dagger}$, $\text{Min Zhang}^{\dagger}$*, $ \text{Yue Cao}^{\ddagger}$, $\text{Woojeong Jin}^{\ddagger}$, $\text{Chenyang Wang}^{\dagger}$,}
\author{$\text{Yiqun Liu}^{\dagger}$, $\text{Shaoping Ma}^{\dagger}$, $\text{Xiang Ren}^{\ddagger}$*}
\affiliation{
  \institution{{$\dagger$} \text{Department of Computer Science and Technology, Institute for Artificial Intelligence}\\ Beijing National Research Center for Information Science and Technology, Tsinghua University, Beijing, China\\
  $\ddagger$ \text{Department of Computer Science, University of Southern California, Los Angeles, CA, USA}\\
  {mawz14@mails.tsinghua.edu.cn, \{z-m, yiqunliu, msp\}@tsinghua.edu.cn, \\ \{cao517, woojeong.jin, xiangren\}@usc.edu, thuwangcy@gmail.com}}
}

% \affiliation{%

% \affiliation{%

% \affiliation{%
 
% \affiliation{%

\begin{abstract}
Explainability and effectiveness are two key aspects for building recommender systems. Prior efforts mostly focus on incorporating side information to achieve better recommendation performance. 
However, these methods have some weaknesses: 
(1) prediction of neural network-based embedding methods are hard to explain and debug;
(2) symbolic, graph-based approaches (e.g., meta path-based models) require manual efforts and domain knowledge to define patterns and rules, and ignore the item association types (e.g. substitutable and complementary).
In this paper, we propose a novel joint learning framework to integrate \textit{induction of explainable rules from knowledge graph} with \textit{construction of a rule-guided neural recommendation model}.
The framework encourages two modules to complement each other in generating effective and explainable recommendation:
1) inductive rules, mined from item-centric knowledge graphs, summarize common multi-hop relational patterns for inferring different item associations and provide human-readable explanation for model prediction; 
2) recommendation module can be augmented by induced rules and thus have better generalization ability dealing with the cold-start issue. 
Extensive experiments\footnote{Code and data can be found at: \url{https://github.com/THUIR/RuleRec}} show that our proposed method has achieved significant improvements in item recommendation over baselines on real-world datasets. Our model demonstrates robust performance over ``noisy" item knowledge graphs, generated by linking item names to related entities.
\end{abstract}

\maketitle

{\fontsize{8pt}{8pt} \selectfont
 \textbf{ACM Reference Format:}\\
 Weizhi Ma, Min Zhang, Yue Cao, Woojeong Jin, Chenyang Wang, Yiqun Liu, Shaoping Ma, Xiang Ren. 2019. Jointly Learning Explainable Rules for Recommendation with Knowledge Graph. In \textit{Proceedings of the 2019 World Wide Web Conference (WWW'19), May 13-17, 2019, San Francisco, CA, USA.} ACM, New York, NY, USA, 11 pages. https://doi.org/10.1145/3308558.3313607}
 
 \vspace{-1em}

\section{Introduction}
Recommender systems play an essential part in improving user experiences on online services. 
While a well-performed recommender system largely reduce human efforts in finding things of interests, often times there may be some recommended items that are unexpected for users and cause confusion. Therefore, explanability becomes critically important for the recommender systems to provide convincing results---this helps to improve the effectiveness, efficiency, persuasiveness, transparency, and user satisfaction of recommender systems \cite{zhang2018explainable}. 

\begin{figure}[t]
    \centering
    \includegraphics[width=0.6\linewidth]{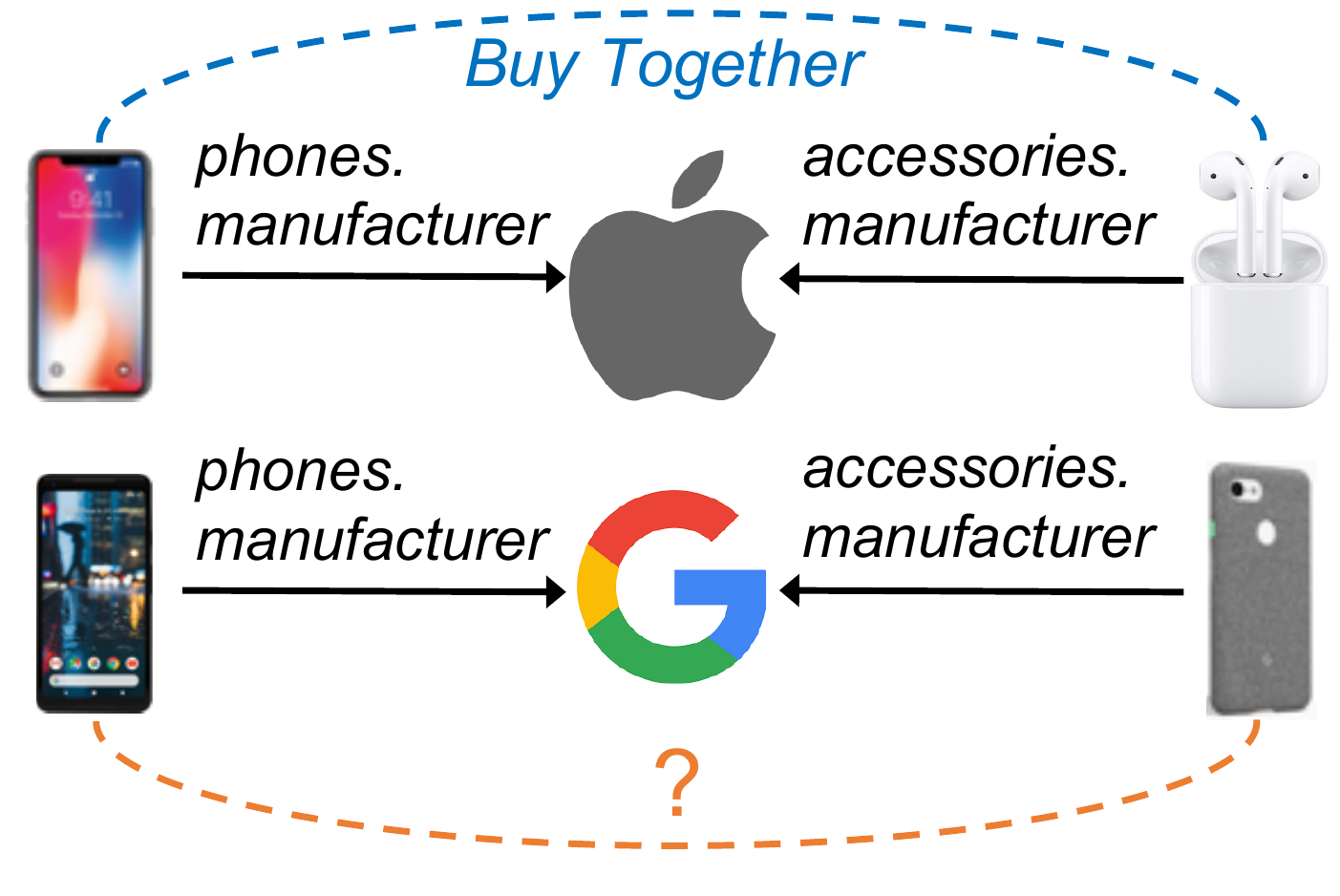}
    \vspace{-0.3cm}
    \caption{\textbf{Illustration of Item-item Associations in a Knowledge Graph.} Given items, relations and item associations (e.g. \textit{Buy Together}), our goal is to induce rules from them and recommend items from rules. These rules are used to infer associations between new items, recommend items, and explain the recommendation.}
    \label{fig:motive}
    \vspace{-1em}
\end{figure}

\makeatletter{\renewcommand*{\@makefnmark}{}
\footnotetext{$^\star$Corresponding author}}

Though there are many powerful neural network-based recommendation algorithms proposed these years,  most of them are unable to give explainable recommendation results \cite{he2017nfm,he2017neural,ma2018your}. Existing explainable recommendation algorithms are mainly two types: user-based \cite{wang2014also, park2017uniwalk} and review-based \cite{zhang2014explicit, he2015trirank}. However, both of them are suffering from data sparsity problem, it is very hard for them to give clear reasons for the recommendation if the item lacks user reviews or the user has no social information. 

On another line of research, some recommendation algorithms try to incorporate knowledge graphs, which contain lots of structured information,  to introduce more features for the recommendation. 
There are two types of works that utilize knowledge graphs to improve recommendation: meta-path based methods~\cite{zhao2017meta,shi2015semantic,yu2013recommendation} and embedding learning-based algorithms~\cite{palumbo2017entity2rec,zhang2016collaborative,shi2018heterogeneous}.
However, 
meta-path based methods require manual efforts and domain knowledge to define patterns and paths for feature extraction. 
Embedding based algorithms use the structure of the knowledge graph to learn users' and items' feature vectors for the recommendation, while the recommendation results are unexplainable. Besides, both types of algorithms ignore item associations.

We find that associations between items/products can be utilized to give accurate and explainable recommendation. For example, if a user buys a cellphone, it makes sense to recommend him/her some cellphone chargers or cases (as they are complementary items of the cellphone). But it may cause negative experiences if the system shows him/her other cellphones immediately (substitute items) because most users will not buy another cellphone right after buying one. So we can use this signal to tell users why we recommend an item for a user with explicit reasons (even for cold items). 
Furthermore, we propose that an idea to make use of item associations: After mapping the items into a knowledge graph, there will be multi-hop relational paths between items. Then, We can summarize explainable rules from for predicting association relationships between each two items and the induced rules will also be helpful for the recommendation.

\begin{figure*}[!t]
    \centering
    \includegraphics[width=\linewidth]{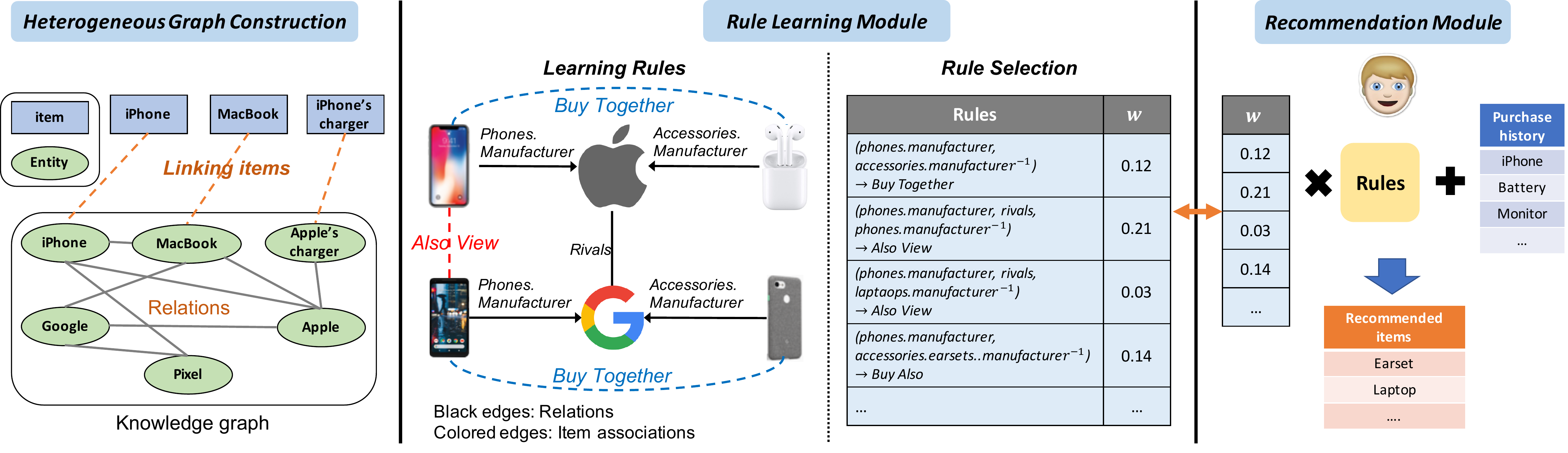}
    \caption{\textbf{Overview of the Proposed RuleRec Framework}. First, we build a heterogeneous graph from items and a knowledge graph. The rule learning module learns the importance of rules and the recommendation module learns the importance at the same time by sharing a parameter vector $\mathbf{w}$.}
    \label{fig:overview}
\end{figure*}

To shed some light on this problem, we propose a novel joint learning framework to give accurate and explainable recommendations.
The framework consists of a rule learning module and a recommendation module.
We exploit knowledge graphs to induce explainable rules from item associations in the rule learning module and provide rule-guided recommendations based on the rules in the recommendation module.
Fig.~\ref{fig:motive} shows an example of items with item associations in a knowledge graph. Note the knowledge graph here is constructed by linking items into a real knowledge graph, but not a heterogeneous graph that only consists of items and their attributes. 
The rule learning module leverage relations in a knowledge graph to summarize common rule patterns from item associations, which is explainable.
The recommendation module combines existing recommendation models with the reduced rules, thus have a better ability to deal with the cold-start problem and give explainable recommendations.
Our proposed framework outperforms baselines on real-world datasets from different domains.
Furthermore, it gives an explainable result with the rules.

Our main contributions are listed as follows:
\begin{itemize}
\item We utilize a large-scale knowledge graph to derive rules between items from item associations. 
\item We propose a joint optimization framework that induces rules from knowledge graphs and recommends items based on the rules at the same time.
\item We conduct extensive experiments on real-world datasets. 
Experimental results prove the effectiveness of our framework in accurate and explainable recommendation
\end{itemize}

\label{intro}

\section{Preliminaries}
% \begin{table}[!t]
%     \centering
%     \caption{Notations.}
%     \begin{tabularx}{\columnwidth}{cl}
%         \toprule
%         \textbf{Notation} & \textbf{Definition} \\
%         \midrule
%         ${U, I_u}$ & {user set, item set for user $u$}\\
%         $R, r$ & {rule, relation (edge) type}\\
%         $E, e$ & {entity, edge}\\
%         $N(a, R)$ & {node set reached with a rule $R$ from node $a$} \\
%         $\mathbf{w}$ & {shared parameter for modules}\\
%         $\mathbf{x}_{(a, b)}$ & {rule feature vector for an item pair $(a,b)$}\\
%         ${F}_{(i, k | R)}$ & {}\\
%         $S_{u,i}, S'_{u,i}$ & {}\\
%         \bottomrule
%     \end{tabularx}
%     \label{tab:notations}
% \end{table}

We firstly introduce concepts and give a formal problem definition. Then, we briefly review BPRMF~\cite{rendle2009bpr} and NCF~\cite{he2017neural} algorithms.
% The notations frequently used in this paper are listed in Table~\ref{tab:notations}.

\subsection{Background and Problem}

\noindent
\textbf{Item recommendation.}
Given users $U$ and items $I$, the task of item recommendation aims to identify items that are most suitable for each user based on historical interactions between users and items (e.g. purchase history).
A user expresses his or her preferences by purchasing or rating items.
These interactions can be represented as a matrix.
One of the promising approaches is a matrix factorization method which embeds users and items into a low dimensional latent space.
This method decomposes the user-item interaction matrix into the product of two lower dimensional rectangular matrices $\mathbf{U}$ and $\mathbf{I}$ for a user and an item, respectively.
From these matrices, we can recommend new items to users.

% Let ${U}$ and ${I}$ denote the set of users and items respectively. 
% Each user $u \in {U}$ has a set of items $I_u = \{i_1, i_2, ... ,i_{T_u}\}$, in which $i_k \in I$ is an item the user $u$ purchased. $T_u$ is the total number of items purchased by user $u$. The goal of recommender system is to recommend a new item list $I'$ to user $u$ based on his/her purchase history $I_u$. 
% ($P(L_u|u, I_u)$).

\smallskip
\noindent
\textbf{Knowledge graph.} A knowledge graph is a multi-relational graph that composed of entities as nodes and relations $r$ as different types edges $e$. 
% Note that multiple edges may belongs to the same relation type.
We can use many triples (head entity $E_1$, relation type $r_1$, tail entity $E_2$) to represent the facts in the knowledge graph \cite{wang2014knowledge}. 

\smallskip
\noindent
\textbf{Inductive rules on knowledge graph.} 
There are several paths between two entities in the knowledge graph, and a path is consisted of entities with the relation types (e.g. $P_k=E_1r_1E_2r_2E_3$ is a path between $E_1$ and $E_3$). A rule $R$ is defined by the relation sequence between two entities, e.g. $R=r_1r_2$ is a rule. The difference between paths and rules is that rules focus on the relation types, not entities. 

\smallskip
\noindent
\textbf{Problem Definition.} 
Our study focus on jointly learning rules in a knowledge graph and a recommender system with the rules.
% in  to boost the recommender systems with derived rules from the knowledge graph. 
Formally, our problem is defined as follows: 
\begin{definition}[Problem Definition]
\textbf{Given} users $U$, items $I$, user-item interactions, item associations, and a knowledge graph, our framework \textbf{aims to} jointly (1) learn rules $R$ between items based on item associations and (2) learn a recommender system to recommend items $I_u'$ to each user $u$ based on the rules $R$ and his/her interaction history $I_u$. This framework outputs a set of rules $R$ and recommended item lists $I'$.
\end{definition}

\subsection{Base Models for Recommendation}

The framework proposed in our study is flexible to work with different recommendation algorithms. As BPRMF is a widely used classical matrix factorization algorithm and NCF is a state-of-the-art neural network based recommendation algorithm, we choose to modify them to verify the effectiveness of our framework.

\smallskip
\noindent
\textbf{Bayesian Personalized Ranking Matrix Factorization (BPRMF).} 
Matrix Factorization based algorithms play a vital role in recommender systems. The idea is to represent each user/item with a vector of latent features. $\mathbf{U}$ and $\mathbf{I}$ are user feature matrix and item feature matrix respectively, and we use $\mathbf{U_u}$  to denote the feature vector of user $u$ ($\mathbf{I_i}$ for item $i$). The dimensions of them are the same. In BPRMF algorithm \cite{rendle2009bpr}, the preference score $S_{u,i}$ between $u$ and $i$ is computed by the inner product of $\mathbf{U_u}$ and  $\mathbf{I_i}$:
\begin{equation}
\label{BPRMF_score}
    S_{u,i} = \mathbf{U}^{\top}_u \cdot \mathbf{I}_i
\end{equation}

The objective function of BPRMF algorithm is defined as a pair-wised function as follows: % \eqref{bprmf_obj},

\begin{equation}
\label{bprmf_obj}
\begin{aligned}
    O_{BPRMF} &= \sum_{u \in U} \sum_{p \in I_u, n \notin I_u} (S_{u,{p}} - S_{u,{n}})
\end{aligned}
\end{equation}
where ${p}$ is a positive item that user $u$ interacted before, and ${n}$ is a negative item sampled randomly from the items user $u$ has never interacted (${n}$ should not be in test set too). 
% we will random sample a  to construct a training pair.

\medskip
\noindent
\textbf{Neural Collaborative Filtering (NCF).}~NCF ~\cite{he2017neural} is a neural based matrix factorization algorithm. Similar to BPRMF, each user $u$ and each item $i$ has a corresponding feature vector $\mathbf{U}_u$ and $\mathbf{I}_i$, respectively. 
% While only the inner product is insufficient to capture some complex user-item interactions, so 
NCF propose a generalized matrix factorization (GMF) (Eq~\eqref{GMF}) and a non-linear interaction part via a multi-layer perception (MLP) (Eq~\eqref{MLP}) between user and item to extraction.
\begin{equation}
\label{GMF}
    \mathbf{h}_{u,i} = \mathbf{U}^{\top}_u \cdot \mathbf{I}_i
\end{equation}
\begin{equation}
\label{MLP}
\begin{split}
\mathbf{g}_{u,i} = \phi_n (...\phi_2(\phi_1(\mathbf{z}_1))) \\
\mathbf{z}_1 = \phi_0 (\mathbf{U}_u \oplus \mathbf{I}_i)\\
\phi_k(\mathbf{z}_{k-1}) = \phi_k (\mathbf{W}^{T}_k\mathbf{z}_{k-1} + \mathbf{b}_{k-1}),
\end{split}
\end{equation}
where $n$ is the number of hidden layers. $\mathbf{W}_k$, $\mathbf{b}_l$, and $\mathbf{z}_k$ are weight matrices, bias vector, and output of each layer. $\oplus$ is vector concatenation and $\phi$ is a non-linear activation function. Both $\mathbf{h}_{u,i}$ and $\mathbf{g}_{u,i}$ are user-item interaction feature vectors for GMF and MLP, respectively. The prediction equation of NCF is defined in Eq~\eqref{NCFpredict}, in which the outputs of GMF and MLP parts are concatentated to get the final score.
% The original objective function of NCF is a point-wised prediction for item rating prediction. 
And we modified the objective function of NCF into Eq~\eqref{objectiveNCF} in this paper.
% study for item recommendation.
\begin{equation}
\label{NCFpredict}
    S_{u,i} = \phi (\alpha \cdot \mathbf{h}_{u,i} \oplus (1-\alpha) \cdot \mathbf{g}_{u,i})
\end{equation}
\begin{equation}
\label{objectiveNCF}
\begin{aligned}
    O_{NCF} &= \sigma(\sum_{u \in U} \sum_{p \in I_u, n \notin I_u} (S_{u,{p}} - S_{u,{n}}))
\end{aligned}
\end{equation}

% \noindent
% \textbf{Rule Learning}: Rule learning is to extract and select a discriminating rule set $R$ according to the rules derived from the entity pairs with a specific association.
% *I think I need your help to refine this definition...*

% \noindent
% \textbf{Recommendation with derived rules}: Let $R$ denote the set of rules derived from the knowledge graph, the recommendation problem changes to $P(L_u|u, I_u, R)$. Considering the rules in the set $R$, we can rewrite the problem as follows:

% \begin{equation}
%     P(L_u|u, I_u, R) = \sum_{r_i \in R} P(I_i|u, I(u), r_i) * P(r_i)
%     \label{prob:reco}
% \end{equation}

% To provide effective recommendation to users with the derived rules, in this paper, we propose to model both rule learning and the recommendation in a multi-task learning framework.

\label{problem}

\section{The \textsf{RuleRec} Framework}
\begin{figure}[t]
    \centering
    \includegraphics[width=1\linewidth]{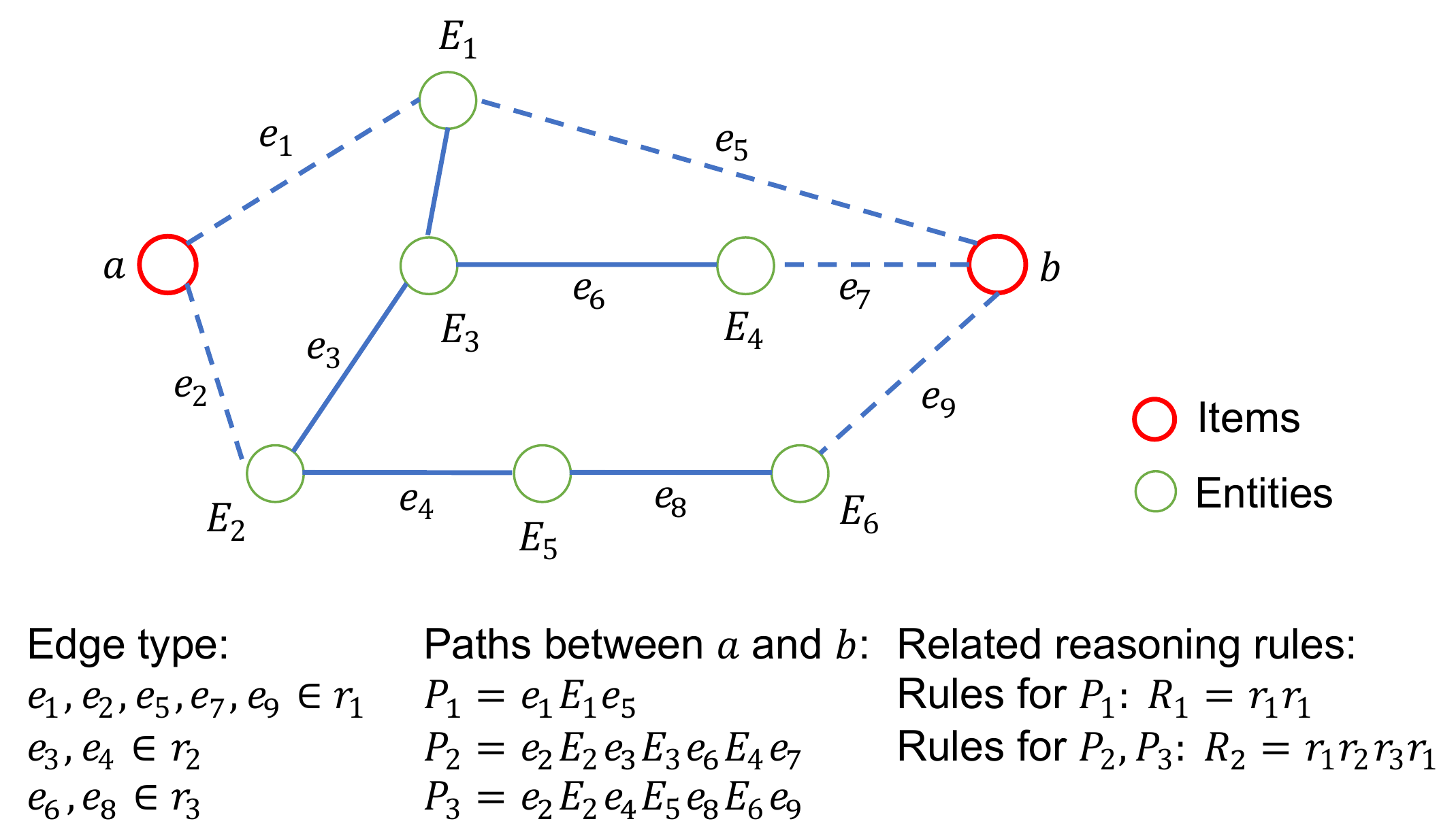}
    \caption{An example of a heterogeneous graph which consists of items and entities in a knowledge graph. The dashed lines are links between items and entities generated by an entity linking algorithm.}
    \label{fig:item}
    % \vspace{-0.5em}
\end{figure}

\noindent
\textbf{Framework Overview.}
Recommendation with rule learning consists of two sub-tasks: 1)~rule learning in a knowledge graph based on item associations; 2)~recommending items for each user $u$ with his/her purchase history $I_u$ and the derived rules $R$.
% We will Introduce our approach that introduce knowledge graph derived rules into the recommendation for explainable recommendation results. There are two sub-tasks: 1) Reasoning rule learning based on item associations; 2) Calculate the recommendation item list for a user u with his/her purchase history $I_u$.

To cope with these tasks, we design a multi-task learning framework. 
The framework consists of two modules, a rule learning module and a recommendation module. 
% Specifically, there are two modules aiming to finish the tasks respectively, namely a rule learning module and a recommendation module. 
The rule learning module aims to derive useful rules through reasoning rules with ground-truth item associations in the knowledge graph. 
Based on the rule set, we can generate an item-pair feature vector whose each entry is an encoded value of each rule. 
The recommendation module takes the item-pair feature vector as additional input
% extra information 
% to refine the original prediction function of recommendation models. 
to enhance recommendation performances and give explanations for the recommendation.
We introduce a shared rule weight vector $\mathbf{w}$ which indicates the importance of each rule in predicting user preference, and shows the effectiveness of each rule in predicting item pair associations. 
% Note that the rule weight in recommendation module indicates the importance of it in prediction user preference, and the rule weight in rule learning module shows the effectiveness of it in selecting useful rules. 
Besides, based on the assume that useful rules perform consistently in both modules with higher weights, 
% and the two sub-tasks are not totally independent. 
% The absolute value of the rule weight indicates the importance of this rule in this module. 
% In both modules, each rule is related with a weight parameter that represents if this rule is important for item association prediction or the recommendation. 
we design a objective function to conduct jointly learning:
% make the two modules a multi-task learning task (equation~\eqref{method:obj}). 
% The overall objective is summarized as follows:
% \begin{equation}
%     \min \limits_{P,W} Loss_{global} = \min \limits_{P,W} \{Loss_{recom} + \beta * Loss_{rule}\} 
% \end{equation}
\begin{equation}
      \min \limits_{V,W} O = \min \limits_{V,W} \{O_{r} + \lambda O_{l}\}
    \label{method:obj}
\end{equation}
where $V$ denotes the parameters of the recommendation module, and $W$ represents the shared parameters of the rule learning and the recommendation module.
The objective function consists of two terms: $O_{r}$ is the objective of the recommendation module, which recommends items based on the induced rules.
$O_{l}$ is the objective of the rule learning module, in which we leverage the given item associations to learn useful rules.
$\lambda$ is a trade-off parameter.
% Fig.~\ref{fig:overview} shows the overall framework.

% In the loss function, 
% $O_{P}$ and $Loss_{rule}$ are the loss function of the two modules respectively, W represents the weight parameters of the derived rules and P represents the other parameters of the recommendation module. Note that in this framework, we aim to make the learning results reasonable, and a good rule should be helpful to the two sub-tasks, the result with only recommendation is not enough. So W is defined as a shared parameter in rule learning module and recommendation module. It will be helpful to make the rule weight W able to capture the importance of the rule in both the recommendation and item association prediction. 
% Next, we introduce the model details.

\subsection{Heterogeneous Graph Construction}
% } 
First, we build a heterogeneous graph containing items for the recommendation and a knowledge graph.
% Specifically, we map the items in recommender systems to the entities in the knowledge graph.
% To copy with this problem, the first step of this module  is to map the items in recommender systems to the entities in the knowledge graph. 
For some items, we can conduct exactly mapping between the item and the entity, such as ``iPhone", ``Macbook". 
For other items, it is hard to find an entity that represents the items, such iPhone's charger. 
Thus, we adopt entity linking algorithm \cite{isem2013daiber} to find the related entities of an item from its title, brand, and description in the shopping website. In this way, we can add new nodes to the knowledge graph that represents items and add some edges for it according to entity linking results.
Then, we get a  heterogeneous graph which contains the items and the original knowledge graph. Fig.~\ref{fig:item} is an example.

\subsection{Rule Learning Module}
\label{sec:rule_module}
% In this module, our goal is to select the most reliable rule set $R_A$ from the knowledge graph with the known item associations $A$. 
The rule learning module aims to find the reliable rule set $R_A$ associated with given item associations $A$ in the heterogeneous graph.

% will be created.

% higher than 0.6
% http://model.dbpedia-spotlight.org/en/annotate
% from dbpedia to freebase
% 4 hops

\smallskip
\noindent
% \subsubsection{
\textbf{Rule learning.}
\begin{figure}
    \centering
    \includegraphics[width=0.6\linewidth]{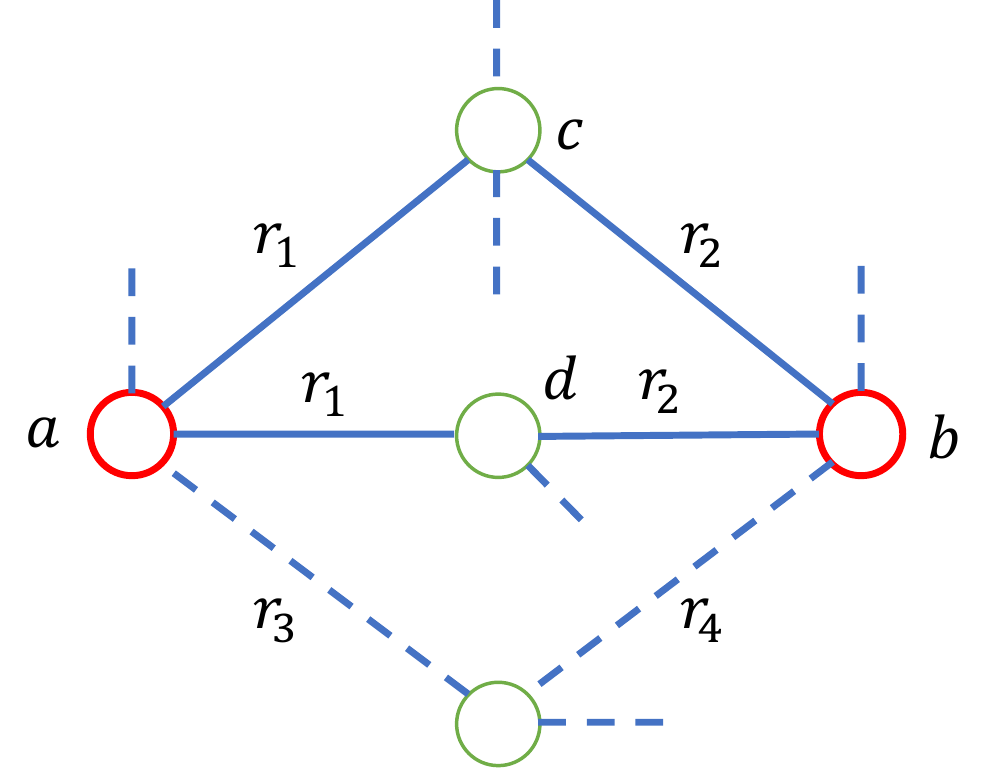}
    \caption{An example of a graph between items $a$ and $b$. $r$ represents a edge type or a relation type.}
    \label{fig:rule}
\end{figure}
% items in a knowledge graph. 
For any item pair ($a$, $b$) in the heterogeneous graph, we use a random walk based algorithm to compute the probabilities of  finding paths which follow certain rules between the item pair, similar to \cite{lao2010relational, lao2011random}. 
Then, we obtain feature vectors for item pairs.
Each entry of the feature vector is the probability of a rule between the item pair.
% rules similar to to follow  find the paths between the item pair \cite{lao2010relational, lao2011random}. 
Here, we focus on relation types between the item pair to obtain rules such as $R_1$ in Fig.~\ref{fig:item},
% (the relation types without entities in a path, such as $r_1r_2r_3$ in Fig.~\ref{fig:item}), 
because it is general to the entities to capture the rules between items.
% easy to adopt the derived rules to calculate the features between another item pair ($i_a$, $I_c$). 

First, we define the probability of a rule between an item pair.
Given a rule $R = r_1...r_k$,
% item $a$ and item $b$, 
% a random walk is defined to calculate the 
probability $P$ with the rule from $a$ to $b$ is defined as:
\begin{equation}
P(b | a, R) = \sum_{e \in N(a, R')} P(e | a, R') \cdot P(b | e, r_k),
\end{equation}
where $R' = r_1...r_{k-1}$, and $P(b|e,r_k)=\frac{I(r_k(e,b))}{\sum_i I(r_k(e,i))}$ is the probability of reaching node $b$ from node $e$ with a one-step random walk with relation $r_k$. 
$I(r_k(e,b))$ is 1 if there exists a link with relation $r_k$ from $e$ to $b$, otherwise 0.
If $b=e$, then $P(b|e,r_k)=1$ for any $r_k$.
% Successor(R', a)  denotes the successor node set of a path from node $a$ with rule $R'$. 
$N(a, R')$  denotes a node set that can be reached with rule $R'$ from node $a$. 
% Generally, 
For example, $P(b|a, R)$ with a rule $R=r_1r_2$ in Fig.~\ref{fig:rule} is computed as follows:
\begin{align*}
P(b | a, R) &= P( c | a, r_1) \cdot P(b | c, r_2) +  P(d | a, r_1) \cdot P(b | d, r_2)
% &=P(c |a, r_1) \cdot P(b | c, r_2) +  P(d |a, r_1) \cdot P(b | d, r_2)
\vspace{-1em}
\end{align*}

Second, we define a feature vector between an item pair.
Given a set of rules, 
% it is easy to calculate 
a rule feature vector for an item pair ($a$, $b$) is defined as $\mathbf{x}_{(a, b)} = [P(b | a, R_1), ... , P(b | a, R_n)]^\top$. 
Each entry in the feature vector $\mathbf{x}_{(a, b)}$ represents a encoded value of rule $R_i$ between $a$ and~$b$. 

% In a large-scale knowledge graph, there are a large number of rules, and it is infeasible to enumerate all possible rules to generate rule feature vectors.
% Thus, we propose rule selection methods in the following section.
% \blue{**How can we select rules?**}

% to select the most useful rules (some of them maybe not reasonable), so we should do some exact analysis.

% \subsubsection{
\smallskip
\noindent
\textbf{Rule selection.}
% \blue{**Is this for selecting rules or computing importance of each rule?**}
To select the most useful rules from the derived rules,  we will introduce two types of  selection methods: hard-selection and soft-selection. 

% we do some further analysis. An intuitively way is to 
\smallskip
\noindent
\textbf{Hard-selection method.} Hard-selection method  set a hyper parameter to decide how many rules we want to select with a selection algorithm firstly. Then we use a chi-square method and a learning based method to choose $n$ rules in this study: 
% \blue{**chi-square method uses weight vector? If not, w is not shared between rule learning module and recommendation module, right?**}
% we use a feature selection method, such as chi-square based and linear regression based algorithms.

% \noindent
(1) Chi-square method. In statistics, the chi-square test is applied to measures dependence between two stochastic variables $A$ and $B$ ~\eqref{chi} (to test if P(AB) = P(A)P(B)).  $N_{A,B}$ is the observed occurrence of two events from a dataset and $E_{A,B}$ is the expected frequency. 
In feature selection, as the features that have lower chi-square scores are independent of prediction target are likely to be useless for classification, chi-square scores between each column of feature vector ($\mathbf{x}_{(a, b)}$) and prediction target ($y_{a,b|A}$) are used to select the top $n$ useful features \cite{schutze2008introduction}.
% with the highest values for the test chi-squared statistic from the prediction target label 
% \blue{A high chi-square value means the entry of feature (probability of a rule in our problem) fits the change of item pair association well, so the top $k$ rules are selected.???}
% , and  and others can be ignored. 
% This is a hard feature filtering method, and the feature vector length will be reduced in this way.
% We use top-$k$ rules based on the chi-square method 
% The number of rule is a hyper-parameter.
\begin{equation}
    \chi^2_{A,B} = \sum \frac{(N_{A,B} - E_{A,B})^2}{E_{A,B}}
    \label{chi}
\end{equation}

% \noindent
% \textbf{Linear regression method.} 
(2) Learning based method. 
% \blue{The number of rule is a pre-defined hyper-parameter in method 1.??} 
Another way to conduct feature selection is to design a objective function $O_{l}$ that compute importance of each rule and try to minimize it.
% with the item pair feature $\mathbf{x}_{(a, b)}$.
In the objective function, we introduce a weight vector $\mathbf{w}$ whose each entry represents importance of each rule.
% The weight of each rule represents how important this feature is. 
For an item pair ($a$, $b$), we use $y_{a,b|A}$ to denote whether $a$ and $b$ have association $A$ ($y_{a,b|A}$ is 1 if they have, and 0 otherwise.). 
We define the following objective functions:

% 2) Linear regression based algorithm use a linear regression model to fit the pair association Y with the rule feature vector $\vec{x}$ and a bias $b$, and the rules with lower weight w will be removed from the feature vector. The final useful rule set of a association A is noted as $R_A$.
% \blue{**$\sum_{all pairs}\sum_{A}??$**}
\begin{itemize}
    \item {Chi-square objective function}
\end{itemize}
\begin{equation}
\label{chi-square:obj}
    \sum_{all pairs \in A} \sum_{i=0}^{|\mathbf{x}_{(a, b)}|} w_i \cdot ( \mathbf{x}_{(a, b)}(i) + b - y_{a,b|A}) ^ 2
\end{equation}
\begin{itemize}
\item {Linear regression objective function}
\end{itemize}
\begin{equation}
\label{linear-regression:obj}
    \sum_{all pairs \in A} \sum_{i=0}^{|\mathbf{x}_{(a, b)}|}  ( w_i \cdot \mathbf{x}_{(a, b)}(i) + b - y_{a,b|A}) ^ 2 
\end{equation}
\begin{itemize}
\item {Sigmoid objective function}
\end{itemize}
\begin{equation}
\label{sigmoid:obj}
    \sum_{all pairs \in A} \sum_{i=0}^{|\mathbf{x}_{(a, b)}|}  \frac{w_i}{1 + e^{-|\mathbf{x}_{(a, b)} (i) + b -y_{a,b|A}|}}  
\end{equation}
% \begin{align}
%     O_{rule} &= \sum_{i=0}^{|X|} W_i * ( X_i + b - Y) ^ 2 \\
%     O_{rule} &=\sum_{i=0}^{|X|}  ( W_i * X_i + b - Y) ^ 2 \\
%     O_{rule} &=\sum_{i=0}^{|X|}  \frac{W_i}{1 + e^{-|X_i + b - Y|}}  
% \end{align}
where $\mathbf{x}_{(a, b)} (i)$ is $i$-th entry of $\mathbf{x}_{(a, b)}$.  To make the objective function reasonable, we constrain that  $\sum_i w_i = 1$ and $w_i > 0$.
In training steps, if $\mathbf{x}_{(a, b)} (i)$ shows positive correlation with $y_{a,b|A}$, then rule $i$ is likely to be useful for item association classification and will get higher weight according to the loss functions. So similar to the chi-square method, the top weighted rules will be selected.  

\smallskip
\noindent
\textbf{Soft-selection method.} Besides the hard-selection method, another way to make use of the learning based objective functions is to take the weight of each rule as a constrain on the rules weights in the recommendation module. No rule will be removed from rule set in this way and it will not introduce extra hyper-parameter. Due to this method is flexible to be combined into other part, 
we introduce the soft-selection method with learning based objective functions to the recommendation module as a multi-task learning. In such condition, there is no extra constrain on rule weight ($\sum_i w_i = 1$ or $w_i > 0$). 
% As item pair with ALV or BAV association are seen as substitute items, the weight should be negative in the recommendation module. So for rules derived from these associations, we change the sign of there loss to make the learning result more reasonable. 
The detail of the multi-task learning method  will be shown in Section 3.5. 

% No te ththe featureat some certain item association between an item pair can be helpful to the recommendation. It sounds like a link prediction problem in the knowledge graph. While different previous studies, our target is not to conduct link prediction (predict whether an item pair has/does not have certain association, e.g.: Also Buy or other) but rule learning in this part. We will calculate the related rules in each association type firstly.

As the rule set is derived from an item association in rule learning module. To apply different item associations at the same time, we can combine the rule sets from different item associations together to get a global rule set $R$.

\subsection{Item Recommendation Module}
\label{sec:reco_module}
We propose a general recommendation module than can be combined with existing methods.
This  module utilizes the derived rule features to enhance recommendation performances.
% In this module, the derived rules are utilized in the recommendation algorithms to enhance performances. 
% The rules are general knowledge about a type of item pair association, so these rules should be helpful to enhance different type of existing recommendation algorithms. Thus, 
% We propose a general framework that can be combined with existing methods.
% is general to  that is flexible to introduce the rule features into different recommendation methods but not just a new algorithm. 

The goal of this module is to predict an item list for user $u$ based on the item set $I_u$ s/he interacted (e.g. purchased) before. 
Previous works calculate the preference score $S_{u,i}$ of user $u$ purchase candidate item $i$, and then rank all candidate items with their scores to get the final recommendation list. 
As shown in Eq~\eqref{score}, we propose a function $f_{\mathbf{w}}$ parameterized by the shared weight vector $\mathbf{w}$ to combine the score $S_{u,i}$ with rule features between candidate item $i$ and items user interacted (e.g. purchased) under rule set $R$.
A score $S'_{u,i}$ for our method is defined as: 
\begin{equation}
    S'_{u,i} = f_{\mathbf{w}}(S_{u,i}, \sum_{k \in I_u} {{F}}_{(i, k | R)})
    \label{score}
\end{equation}

The feature vector for item pair ($a, b$) under rule set $R$ is denoted by  ${F}_{(a,b|R)}$. Note that ${F}_{(a,b|R)}$ is different from $\mathbf{x}_{(a, b)}$ and calculated by ${F}_{(a,b|R)} = \sum_{e \in N(a, R')} P(e | a, R') \cdot I(b | e, r_k)$. $I(b | e, r_k)$ is an indicator function: if there is a edge in relation type $r_k$   between $b$ and $e$, $I(b | e, r_k) = 1$; otherwise 0. 
The reason why we adopt another feature generation method is that in recommendation module, we concerns more about if there exists a path in this rule between two items. The weight of each rule will be used in explaining the recommendation result, so we should make the comparing between rules fair. While longer rules are more likely to get lower score (more random walk steps so lower probability). If the feature vector is still $\mathbf{x}$, it will hurt the explainable of our module. Thus we use ${F}_{(a,b|R)}$ as the feature vector here, which represents the frequency of each rule between the two items.

To consider the global item associations between candidate item $i$ and the item set $I_u$,  we add the rule features between $i$ and each item $I_k$ in $I_u$ together. For convenience, the new feature vector is named as ${F}_{(i, I_u|R)}$. So Eq~\eqref{score} can be rewrite as the following:

\begin{equation}
    S'_{u,i} = f_{\mathbf{w}}(S_{u,i}, {{F}}_{(i, I_u | R)})
    \label{score_re}
\end{equation}

% Write the below sentence in the experiment section.
% There are many candidate functions for $f(a,\mathbf{x})$, such as $f(a,\mathbf{x}) = a + b  (\mathbf{w}^{\top}\mathbf{x})$ and $f(a,\mathbf{x}) = a  (\mathbf{w}^\top  \mathbf{x})$. We adopt different combination methods and the analysis are shown in Section 4.x. 

We define the objective function for the recommendation module as follows:
% Specially, we will modify the original objective function and train the weight vector $\mathbf{w}$ the original recommendation algorithm together. 
% For example, the primary objective of BPRMF algorithm \cite{rendle2009bpr} is a pair-wised loss shown in equation~\eqref{objective}. $\mathbf{u}$ is user embedding  $\mathbf{i}$ maximum the score of positive item and minimize the score of negative item). If the rule features are introduced with $f(a,\mathbf{x}) = a + b (\mathbf{w}^{\top}\mathbf{x})$, the modified objective function is shown in equation~\eqref{objective_new}. The weight vector of the rules are arguments in recommendation model learning too.
\begin{equation}
\label{objective_rule}
\begin{aligned}
    O_{r} &=\sum_{u \in U} \sum_{p \in I_u, n \notin I_u} (S'_{u,p} - S'_{u,n})\\
    &= \sum_{u \in U} \sum_{p \in I_u, n \notin I_u} \left(f_{\mathbf{w}}(S_{u,p}, {{F}}_{(p, I_u | R)}) - f_{\mathbf{w}}(S_{u,n}, {{F}}_{(n, I_u | R)})\right),
\end{aligned}
\end{equation}

% \begin{equation}
%     Loss_{BPRMF} = \sigma(\vec{U_u}^\top (\vec{I_{posi}} - \vec{I_{nega}})
%     \label{loss}
% \end{equation}

% \begin{equation}
% \begin{split}
% % \setlength{\itemsep}{0pt}
% % \setlength{\parsep}{0pt}
% % \setlength{\parskip}{0pt}
%     O_P = \sigma\left(\vec{U_u}^\top (\vec{I_{posi}} - \vec{I_{nega}} + \\ \alpha * \vec{w}^{T} * (\vec{F}_{(i_{posi}, I_u | R)} - \vec{F}_{(i_{nega}, I_u | R)}))\right)
% \end{split}
% \label{loss_new}
% \end{equation}

% \begin{equation}
% \label{objective}
% \begin{aligned}
%     O &= S_{u,i_{posi}} - S_{u,i_{nega}}\\
%     % &=\vec{U_u}^\top * \vec{I_{posi}} - \vec{U_u}^\top * \vec{I_{nega}}) \\
%     % &=\vec{U_u}^\top (\vec{I_{posi}} - \vec{I_{nega}})
% \end{aligned}
% \end{equation}

% \begin{equation}
% \begin{aligned}
%     O_p &= \sum_{p \in P} \sum_{n \in N} S'_{u,i_{posi}} - S'_{u,i_{nega}} \\
%     &= S_{u,i_{posi}} - S_{u,i_{nega}} + \alpha   \vec{w}^{T}  (\vec{F}_{(i_{posi}, I_u | R)} - \vec{F}_{(i_{nega}, I_u | R)})\\
%     &= O +  \alpha  \vec{w}^{T}  (\vec{F}_{(i_{posi}, I_u | R)} - \vec{F}_{(i_{nega}, I_u | R)})
% \end{aligned}
% \label{objective_new}
% \end{equation}

where $p$ is a positive item ($\in I_u$) and $n$ is a random sampled negative item ($\notin I_u$)
% to construct a training pair 
for user $u$.
% Note that the rule weight vector $\mathbf{w}$ is a very important feature vector, because it records the contribution of the corresponding rules for the recommendation results. 
Note that the rule weight vector $\mathbf{w}$ gives explanations for item pairs with rules in recommendation module.
% If the weight of a derived rule is positive, it means item pairs with this rule have higher probability to be complementary products (buy one and more likely to buy the other). If the weight of a rule is negative, it indicates that item pairs with this rule have higher probability to be substitute items (buy one and will not buy the other). 
If a candidate item $i$ gets a higher score than other candidate items, the rule which contributes the highest score for $i$ and the corresponding items the user bought can be used to explain why the algorithm recommends $i$ to the user. In other words, the introduction of rule features make the recommendation results explainable. There are some case studies in Section~5.5.
% and find the important rules.

\begin{figure}
    \centering
    \includegraphics[width=0.75\linewidth]{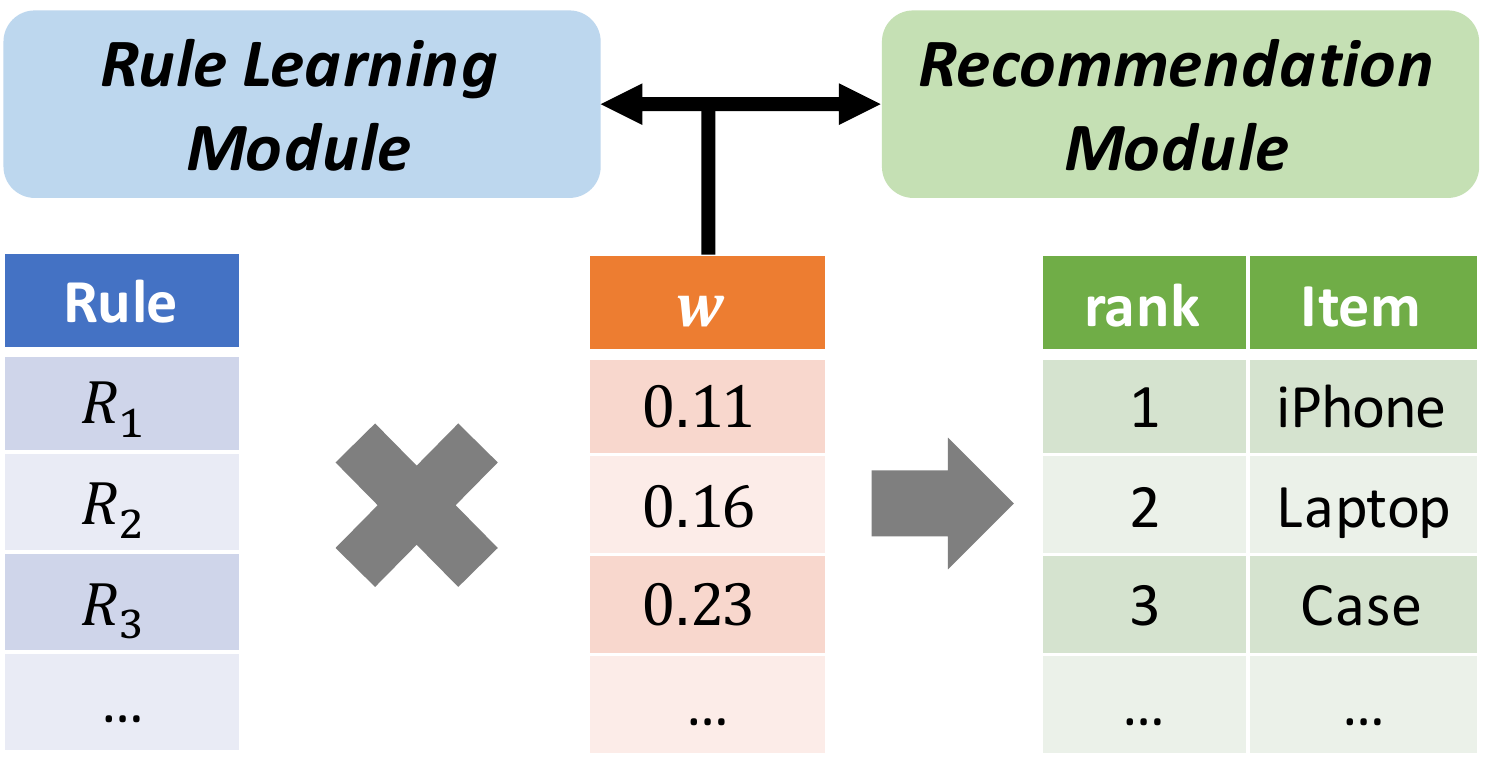}
    \caption{Multi-task learning of the rule learning module and the recommendation module. These two modules share the parameter $\mathbf{w}$.}
    \label{fig:joint}
\end{figure}

This combination method is flexible and easy to introduce rule features to many previous recommendation models (use the algorithm's prediction function to calculate $S_{u,i}$). 
In this study, we implement this recommendation module with BPRMF (traditional MF algorithm) and NCF (neural network based MF algorithm). 
Since it is a two step algorithm (to learn rules firstly and then conduct recommendation), we denote them as RuleRec$_{two}$(BPRMF) and RuleRec$_{two}$(NCF). The prediction function and objective function of them are Eq \eqref{score_re} and Eq \eqref{objective_rule}, where $S_{u,i}$ is replaced by the prediction function of BPRMF (Eq \eqref{BPRMF_score}) and NCF (Eq \eqref{NCFpredict}), respectively.

\begin{table*}[!htbp]
\newcommand{\tabincell}[2]{\begin{tabular}{@{}#1@{}}#2\end{tabular}}
\caption{\textbf{The statistics of item association pairs in different domain.} \#Involved item means the number of items that have at least one type of association with any other items. \#Item is the number of involved items with association and \#Pair is the number of item pairs with association.}
\begin{tabular}{|c|c|c|c|r|c|r|c|r|c|r|}
\hline
\multirow{2}{*}{\textbf{Dataset}} & \multirow{2}{*}{\tabincell{c}{\textbf{\#Item}}} & \multirow{2}{*}{\tabincell{c}{\textbf{\#Involved} \\ \textbf{Item}}} & \multicolumn{2}{c|}{\textbf{Also View}}                              & \multicolumn{2}{c|}{\textbf{Buy After Viewing}}                      & \multicolumn{2}{c|}{\textbf{Also Buy}}                               & \multicolumn{2}{c|}{\textbf{Buy Together}}                         \\ \cline{4-11} 
                                  &                       &                    & \textbf{\#item}               & \multicolumn{1}{c|}{\textbf{\#Pair}} & \textbf{\#item}               & \multicolumn{1}{c|}{\textbf{\#Pair}} & \textbf{\#item}               & \multicolumn{1}{c|}{\textbf{\#Pair}} & \textbf{\#item}              & \multicolumn{1}{c|}{\textbf{\#Pair}} \\ \hline
\textbf{Cellphone}     &     \multicolumn{1}{r|}{346,793}      & \multicolumn{1}{r|}{214,692}             & \multicolumn{1}{r|}{103,845} & 1,038,090                         & \multicolumn{1}{r|}{181,935} & 1,818,990                             & \multicolumn{1}{r|}{71,660} & 716,240                          & \multicolumn{1}{r|}{29,372} & 293,360                             \\ \hline
\textbf{Electronic}   &     \multicolumn{1}{r|}{498,196}      & \multicolumn{1}{r|}{318,922}                     & \multicolumn{1}{r|}{123,959}         &  1,239,230                                    & \multicolumn{1}{r|}{250,409}         &  2,503,730                                    & \multicolumn{1}{r|}{159,562}         &         1,595,260                             & \multicolumn{1}{r|}{31,040}        &  310,040                                    \\ \hline
\end{tabular}
\label{fig:item_assoc}
\vspace{0.0cm}
\end{table*}

\subsection{Multi-task Learning}
In Sections~\ref{sec:rule_module} and \ref{sec:reco_module}, we introduced the two modules respectively. 
We can train the modules one by one 
% use the two step algorithms one by one 
to get the recommendation results. %While 
The shortcoming of training two modules separately is that the usefulness of rules in prediction item association is ignored. 
Instead, we share the rule weight $\mathbf{w}$, and this weight can capture the importance of the rule in both the recommendation and item association prediction simultaneously as shown in Fig.~\ref{fig:joint}.
% Based on the idea that make the rule weight $\mathbf{w}$ shared will be helpful to capture the importance of the rule in both the recommendation and item association prediction simultaneously, 
Thus, we propose a multi-task learning objective function defined as follows:
\begin{equation}
    O =  O_{r} + \lambda O_{l}
    \label{method:obj2}
\end{equation}
% Both rule learning loss and recommendation module loss are taken into consideration. 
where $O_{l}$ and $O_{r}$ are the objective functions for the rule learning module and the recommendation module, respectively.
% is the objective function for the rule learning module,
% % one of the three rule selection with parameter loss in Section~\ref{sec:rule_module} 
% and $O_{reco}$ is the objective function for the recommendation module
% % the enhanced recommendation algorithm.
% % E.g: Equation 9 for BPRMF algorithm). As $\mathbf{w}$ is the shared parameters for rule learning module and recommendation module, both objective functions will effect the value of it.
Note that both objective functions share $\mathbf{w}$. 

The multi-task learning combination method is able to conduct rule selection and recommendation model learning together. Similar to the two step combination method, it is also flexible to to multiple recommendation models too. BPRMF and NCF are enhanced with this idea, and the modified algorithms are named as RuleRec$_{multi}$(BPRMF) and RuleRec$_{multi}$(NCF).

% With the above settings, both modules can tightly interact with each other in training, and provide extra supervision for parameter learning to finish the two sub-tasks together without extra hyper-parameter. 

\label{proposed}

\section{Rule Selection Details}
This section introduces the implementation details and results of the rule selection component in RuleRec. 

\subsection{Dataset and Implementation Details}
We introduce item association datasets, a knowledge graph, and recommendation datasets for experiments. 

\smallskip
\noindent
\textbf{Item association datasets}. A open dataset with item associations is used in our experiments\footnote{http://jmcauley.ucsd.edu/data/amazon/}. The item associations are extracted from user log on Amazon (same as \cite{mcauley2015inferring}). 
Four types of item associations are considered: 1) Also view (ALV), users who viewed x also viewed y; 2) Buy after view (BAV), users who viewed x eventually bought y; 3) Also buy (ALB), users who bought x also bought y; 4) Buy together (BT), users frequently bought x and y together. ALV and BAV are substitute associations, and ALB and BT are complementary associations. 
The statistics of Cellphone and Electronics datasets with different item associations are shown in Table~1. Since the data is crawled from Amazon\footnote{www.amazon.com}, the number of link is nearly ten times as large as the number of involved items in each association type. Besides, as shown in the table, over 37\% items do not have any association with other items in this dataset.

% \begin{table}[]
% \caption{The statistics of item associations.}
% \begin{tabular}{|c|c|c|c|}
% \hline
% \textbf{Category}& \textbf{Cellphone} & \textbf{Electronics}  \\ \hline
% \textbf{\# items}       & 346, 793  & 498, 196  \\ \hline
% \textbf{Also View}      & 133, 218  & 150, 135  \\ \hline
% \textbf{Buy After View} & 216, 811  & 305, 007  \\ \hline
% \textbf{Also Buy}       & 135, 367  & 257, 975  \\ \hline
% \textbf{Buy together}   & 88, 280   & 113, 435 \\ \hline
% \textbf{\# at least one} & 240, 591 & 353, 695 \\ \hline
% \end{tabular}
% \label{tab:item_assoc}
% \end{table}

\smallskip
\noindent
\textbf{Knowledge graph dataset}. Freebase \cite{bollacker2008freebase} is used to learn rules.
% as a knowledge graph for rule learning here. 
It is the largest open knowledge graph\footnote{https://developers.google.com/freebase/}, containing more than 224M entities, 784K relation types, and over 1.9 billion links. 

% \noindent
% \textbf{Recommendation datasets}. The recommendation datasets are open datasets that extracted from Amazon \cite{mcauley2015image, he2016ups}.  User' purchase histories in Amazon are recorded with the purchased items and times. We conduct experiments using two datasets: Amazon Cellphone and Amazon Electronic. Each user has at least 5 interactions with items. The statistics of the datasets are summarized in Table~\ref{tab:reco_data}. \blue{what is sparsity in the table?}

% \begin{table}[!htbp]
% \caption{The statistics of recommendation datasets.}
% \begin{tabular}{|c|r|r|r|r|}
% \hline
% \textbf{Dataset}    & \multicolumn{1}{c|}{\textbf{\#user}} & \multicolumn{1}{c|}{\textbf{\#item}} & \multicolumn{1}{c|}{\textbf{\#links}} & \multicolumn{1}{c|}{\textbf{Sparsity}} \\ \hline
% \textbf{Cellphone}  & 27, 879                     & 10, 429                     & 194, 439                     & 99.93\%                       \\ \hline
% \textbf{Electronic} & 22, 675                     & 58, 741                     &                              &                               \\ \hline
% \end{tabular}
% \label{tab:reco_data}
% \end{table}

% As introduced in Section 3.2, 
The link prediction algorithm~\footnote{http://model.dbpedia-spotlight.org/en/annotate} \cite{isem2013daiber}  is used to connect items (with their titles, brands, and descriptions) and entities in DBPedia~\footnote{https://wiki.dbpedia.org/} firstly. 
Then the linked entities in DBPedia are mapped to the entities in Freebase with a entity dictionary~\footnote{https://drive.google.com/file/d/0Bw2KHcvHhx-gQ2RJVVJLSHJGYlk/view}. 
% What is the link prediction algorithm? is this for mapping items to a KG? and I can't access the page.
As there is a probability score of each linked entity with the algorithm, which represents the confidence of this linking. So if the probability of a word links to a entity is lower than 0.6, we will ignore it to make the link result more accurate.
% Although this link prediction algorithm is designed for DBPedia, the entities are mapped to Freebase. (*** to do, The number is 3.6 million)

Due to the large scale of the knowledge graph, it is infeasible to enumerate all possible rules in this step. 
Following the idea in \cite{lao2011random}, we require that all derived rule needs to be supported by at least a fraction $\alpha$ of the training item pairs, as well as being of length no more than $\beta$ (there will be huge number of rules without the length constraint). 
In the experiments, we set $\alpha$ to 0.01 (the same as \cite{lao2011random}),  
% Previous studies offer define $\beta$ as 3 or 4, we prefer to mine more rules so 
and $\beta$ to 4, which means the maximum number of edges between entities in a path is 4.

\subsection{Results of Rule Selection}

\textbf{Item linking to the Knowledge Graph.}
In this step, we link the items from different domains to the entities in the knowledge graph.
% The items from different domains are linked with the entities in this step. 
Items in the Cellphone domain and the Electronic domain are connected with 33,542 entities and 55,180 entities in Freebase respectively. Due to the item-entity linking method is not in a one-by-one accurate linking but based on items' titles, brands, and descriptions, each item will be linked into several entities and each entity will be linked with several items. 
With the random walk strategy introduced in Section~\ref{sec:rule_module}, we find that the four hop routes in the knowledge graph from these entities will pass over 10 million entities. To avoid introducing unrelated entities in random walk step, the type of entities are constrained on pre-defined entity types (e.g.: entities in ``ns.base.brand", ``ns.computer'' and some other types are maintain), then the involved entity amount is reduced to around 100K in each domain.

% link as much as items into the knowledge graph with entity linking, while it is still impossible to link all items to it (some items lack title and brand information in the dataset). The link result of 

% \begin{table}[]
% \begin{tabular}{|c|r|r|r|r|r|}
% \hline
% \textbf{Dataset}     & \multicolumn{1}{c|}{\textbf{\#ALV}} & \multicolumn{1}{c|}{\textbf{\#BAV}} & \multicolumn{1}{c|}{\textbf{\#ALB}} & \multicolumn{1}{c|}{\textbf{\#BT}} & \multicolumn{1}{c|}{\textbf{\#Item}} \\ \hline
% \textbf{Cellphone}   & 133, 218                            & 216, 811                            & 135, 367                            & 88, 280                            & 346, 793                             \\ \hline
% \textbf{Electronics} & 150, 135                            & 305, 007                            & 257, 975                            & 113, 435                           & 498, 196                             \\ \hline
% \end{tabular}
% \end{table}

\smallskip
\noindent
\textbf{Rule Learning.} 
% Based on the rule learning setting in Section 4.1 (no more than 4 hops and supported by at least 0.01 training item pairs), 
The derived rules of different associations in cellphone domain are summarized in Table~\ref{tab:deriv_rules}. 
% Due to the sparsity of each domain and association is distinct, the numbers of derived rules are quiet different. 
There are hundreds of rules derived from Cellphone domain in each association, while only around 46-70 rules are in Electronic domain.
The possible reason is that comparing with Cellphone domain, Electronic domain contains more items and the items are more diversity. Most rules are supported by less than 0.01 of the training item pairs. so less general rules are derived.

\begin{table}[]
\label{tablerulelearning}
\caption{The number of derived rules from different associations.}
\begin{tabular}{|c|r|r|r|r|}
\hline
\textbf{Dataset}     & \multicolumn{1}{c|}{\textbf{\#ALV}} & \multicolumn{1}{c|}{\textbf{\#BAV}} & \multicolumn{1}{c|}{\textbf{\#ALB}} & \multicolumn{1}{c|}{\textbf{\#BT}}  \\ \hline
\textbf{Cellphone}   & 700                            & 948                            & 735                            & 675                                                       \\ \hline
\textbf{Electronic}   & 46                            & 66                            & 70                            & 50                                                       \\ \hline
\end{tabular}
\label{tab:deriv_rules}
\end{table}

\smallskip
\noindent
\textbf{Rule Selection.} 
To select useful rules from the large rule set, we use the learning based (LR, Eq~(11)) and chi-square based feature selection methods in Section~\ref{sec:rule_module}. 
The idea of selection methods is to choose the rules by which any items in a specific association are followed.
% that perform good in predicting whether an item pair is in a specific association. 
% In other words, if there is a rule $r_k$ between any item pair in BT association, and none when an item pair is not in BT association, $r_k$ is a perfect rule.
E.g. if any item pairs in the BT association follows a rule $R_k$, then $R_k$ is a useful rule for the BT association.

We choose the ALB association in the Cellphone dataset to verify the selection ability of the two methods.
Because the derived rules will be used to extract item-item pair feature for the recommendation, a good rule should be able to indicate the associations between item $i$ and user's purchase history $I_u$. So the recommendation dataset (Section 5.1) in the Cellphone domain is used for evaluation, we calculate the recall of whether there is at least one path satisfied rule $r_k$ between the last item $i_l$ user interacted and user's previous purchase history $I_u$. 
Due to not always exist at least one rule between $i_l$ and $I_u$,  there is a upper bound for the recall. 

\begin{table}[]
\centering
\caption{Rule selection results on ALB association in the Cellphone domain. }
\begin{tabular}{|c|c|c|c|c|c|}
\hline
 \multicolumn{2}{|c|}{\textbf{LR}} & \multicolumn{2}{c|}{\textbf{Chi-square}} & \multirow{2}{*}{\textbf{All}} & \multirow{2}{*}{\begin{tabular}[c]{@{}c@{}}\textbf{Upper} \\ \textbf{Bound}\end{tabular}} \\ \cline{1-4}
                                                                          Top 50     & Top 100    & Top 50         & Top 100        &                      &                                                                         \\ \hline
20.1\%     & 40.1\%     & 87.0\%         & 88.5\%         & 89.2\%               & 90.7\%                                                                  \\ \hline
\end{tabular}
\label{fig:rule_sel}
\end{table}

Table~\ref{fig:rule_sel} shows the rule selection results ALB association in Cellphone domain and its upper bound. 
Chi-square based method outperforms linear-regression based method in rule selection. The reason is that rules with higher weight in linear regression model cannot fully represent usefulness of rules in the recommendation.  However, Chi-square method is able to find the most useful rules, and the selected 50 rules cover 87.0\% of user purchase history (only 2.2\% percentage lower than using all rules). 
It is reasonable to choose only the subset of derived rules for the recommendation. % in two-step model. 
Besides, we find that the upper bound in Electronic domain is only about 65\%, indicating that the combination between rules in Electronic dataset is not as tightly as in the Cellphone dataset.

Besides, for multi-task learning framework, it is unnecessary to conduct rule selection because the model takes the effect of each rule in predicting item associations through the combined loss function Eq  (\eqref{chi-square:obj}, \eqref{linear-regression:obj}, or \eqref{sigmoid:obj}).

% \usepackage{multirow}

% \usepackage{multirow}

\label{rule}

\section{Recommendation Experiments}
% \usepackage{multirow}

% \usepackage{multirow}

This section introduces dataset and experiment settings for comparing RuleRec with other baseline methods, as well as providing case study on analyzing different components of RuleRec.

\subsection{Recommendation Dataset}
The recommendation datasets are open datasets that extracted from Amazon \cite{mcauley2015image, he2016ups}.  Each user's purchase histories in Amazon are recorded with the purchased items and times. We conduct experiments using two datasets: Amazon Cellphone and Amazon Electronic. Each user has at least 5 interactions with items.  The statistics of the datasets are summarized in Table~\ref{tab:reco_data}.

\begin{table}[!htbp]
\caption{The statistics of recommendation datasets.}
\vspace{-0.2cm}
\begin{tabular}{|c|r|r|r|}
\hline
\textbf{Dataset}    & \multicolumn{1}{c|}{\textbf{\#user}} & \multicolumn{1}{c|}{\textbf{\#item}} & \multicolumn{1}{c|}{\textbf{\#links}}  \\ \hline
\textbf{Cellphone}  & 27, 879                     & 10, 429                     & 194, 439                                          \\ \hline
\textbf{Electronic} & 22, 675                     & 58, 741                     & 195, 751                                                          \\ \hline
\end{tabular}
\label{tab:reco_data}
\end{table}
% The basic statistics of the recommendation datasets has been introduce in Section 4.1, recommendation datasets part. We conduct experiments using two datasets: Amazon Cellphone and Amazon Electronic. 

\subsection{Experimental Settings}

\smallskip
\noindent
\textbf{Evaluation Protocol.} To evaluate the item recommendation performance, we use leave-one-out evaluation in the recommendation~\cite{bayer2017generic, he2017neural}. The latest interactions between items and each user are used as positive items in  test set, and the remaining data are used for training. 
Due to the loss function in our study is pair-wised, each positive item in training set will be trained with a negative item sampled from items that the user has not interacted. 
As for test set, since it is too time-consuming to rank all items for each user in evaluation, 99 negative items that are not interacted with the user are random sampled and added to test set \cite{cheng2018delf, wang2018matrix}. Therefore, in the test set, each user is evaluated with 99 negative items and one positive item. The target here is to generate a high-quality ranked list of items for each user. 

\smallskip
\noindent
\textbf{\textbf{Evaluation Metrics.}} We use Recall, Normalized Discounted Cumulative Gain (NDCG), and Mean reciprocal rank (MRR). Higher score means better performance in each metric. Recall focuses on whether the positive item is in the list, while NDCG and MRR take the position of the positive item into evaluation. 
Considering that the length of most recommendation list in real scenarios is 5 or 10, so the ranked list is truncated at 10 for all metrics. We calculate Recall@5, Recall@10, NDCG@10, and MRR@10 for evaluation.  

\subsection{Compared Methods}
Three types of baselines (traditional matrix factorization, neural network based, and recommendation with knowledge graph) are used here:
\begin{itemize}[leftmargin=0.4cm,noitemsep,nolistsep]
    \item \textbf{BPRMF} \cite{rendle2009bpr}. As introduced in Section 2.2.1, this method follows the idea of matrix factorization with  pairwise ranking loss. 
    \item \textbf{NCF} \cite{he2017neural}: This is a state-of-the-art latent factor model. It pre-trains MLP and GMF part separately, and then ensembles the two models to get the final preference score. Following previous studies \cite{cheng2018delf, he2018outer}, \textbf{MLP} and \textbf{GMF} are taken as baseline models too.
    % The generalized matrix factorization part of NCF algorithm. 
    % The multi-layer perception part of NCF algorithm, user vector and item vector are concatenated as a new vector, and the new vector is  inputted into a multi-layer perception for preference prediction. 
    \item \textbf{HERec} \cite{shi2018heterogeneous}: A state-of-the-art algorithm which using the knowledge graph for the recommendation. This method adopts meta-paths to generate the embeddings of users and items in the heterogeneous network with Deepwalk~\cite{Perozzi2014DeepWalkOL}, and then use them in the recommendation. Two variants of this algorithm with different fusion functions, \textbf{HERec$_{sl}$} (with the simple linear fusion function) and \textbf{HERec$_{pl}$} (with personalized linear fusion function)  are used as baseline models.
    \item \textbf{RippleNet} \cite{wang2018ripplenet}: Another state-of-the art algorithm that  incorporates the knowledge graph into recommender systems. It stimulates the propagation of user preferences on the set of knowledge entities to learn a user's potential interests.
\end{itemize}

\medskip
\noindent
{\textbf{Implementation Details.}}
We adopt the implementation of BPRMF algorithm in MyMediaLite\footnote{http://www.mymedialite.net/index.html} (a famous open source package) on our experiments. The implementation of other algorithms are from the public codes that the authors provided in their papers (NCF\footnote{https://github.com/hexiangnan/neural\_collaborative\_filtering},  HERec\footnote{https://github.com/librahu/HERec}, and RippleNet\footnote{https://github.com/hwwang55/RippleNet}).  The four new models, {RuleRec$_{two}$} with BPRMF, {RuleRec$_{two}$} with NCF, {RuleRec$_{multi}$} with BPRMF, and {RuleRec$_{multi}$} with NCF are modified from BPRMF and NCF according to our framework respectively. We tune all the parameters to achieve the best performance of each algorithm.

% The combination score function $f_{w}$ 
The score function is defined as $S^{'}_{u,i} = f_\mathbf{w}(S_{u,i}, {{F}}_{(i, I_u | R)}) = S_{u,i} + \alpha \cdot \mathbf{w}^T{{F}}_{(i, I_u | R)}$ in this section. 
Different implementations of $f_\mathbf{w}$ and their results will be analyzed in Section~\ref{sec:comb}. All of the four types of item associations are used in the recommendation module for both two-step and multi-task learning algorithms. Top 50 rules of each type of item associations (selected with chi-square method) are chose to the two-step based methods. To make the comparison fair, 
these rules are used in the multi-task learning algorithms with the sigmoid objective function in the final experiments. The objective function is sigmoid (Eq \eqref{sigmoid:obj}), as it performs the best in the three objective functions (Eq \eqref{chi-square:obj}, \eqref{linear-regression:obj}, and \eqref{sigmoid:obj}); due to the limited of length, we do not show the results here. 
The comparison of different amounts of rules will be introduced in Section~5.5.5.

% \vspace{-2em}

% Two-step v.s. Multi-task learning

\begin{table*}[]
\caption{\textbf{Performance Comparison between RuleRec and Other Methods in Different Domains.} RuleRec$_{two}$ and RuleRec$_{multi}$ are our proposed models. RuleRec$_{two}$ is a two-step rule-based model and RuleRec$_{multi}$ is a multi-task model. These models use BPRMF or NCF as a recommendation model.   * indicates statistical significance at $p < 0.01$ compared to the best baseline model.}
\begin{small}
\begin{tabular}{c|rrrr|rrrr}
\hline
\multirow{2}{*}{\textbf{Methods}~/~\textbf{Dataset}} & \multicolumn{4}{c|}{\textbf{Cellphone}}                                                                                                                         & \multicolumn{4}{c}{\textbf{Electronic}}                                                                                                                        \\ %\cline{2-9} 
                                  & \multicolumn{1}{c}{\textbf{Recall@5}} & \multicolumn{1}{c}{\textbf{Recall@10}} & \multicolumn{1}{c}{\textbf{NDCG@10}} & \multicolumn{1}{c|}{\textbf{MRR@10}} & \multicolumn{1}{c}{\textbf{Recall@5}} & \multicolumn{1}{c}{\textbf{Recall@10}} & \multicolumn{1}{c}{\textbf{NDCG@10}} & \multicolumn{1}{c}{\textbf{MRR@10}} \\ \hline\hline
\textbf{BPRMF}~\cite{rendle2009bpr}                    & 0.3238                                 & 0.4491                                  & 0.2639                                & 0.2058                               & 0.1886                                 & 0.2763                                  & 0.1571                                & 0.1207                               \\ %\hline
\textbf{GMF}~\cite{he2018outer}                      &           0.3379                            &                                 0.4666       &                  0.2789                      &      0.2223                               &             0.1988                            &    0.2835                                     &            0.1657                           &            0.1298                          \\ %\hline
\textbf{MLP}~\cite{cheng2018delf}           &    0.3374                                    &    0.4779                                     &         0.2790                              &      0.2182            &                0.2000                        &        0.2883                                 &       0.1681                                &            0.1315                                                         \\% \hline
\textbf{NCF}~\cite{he2017neural}                      & 0.3388                                 & 0.4751                                  & 0.2761                                & 0.2151                               &     0.2005                                   &                0.2916                       &                              0.1679         &              0.1300                        \\ \hline
\textbf{Hec$_{sl}$}~\cite{shi2018heterogeneous}                   & 0.2436                                 & 0.3481                                  & 0.2040                                & 0.1600                               & 0.1870                                 & 0.2851                                  & 0.1534                                & 0.1135                               \\ %\hline
\textbf{Hec$_{pl}$}~\cite{shi2018heterogeneous}    &       0.2511        &             0.3564                                                                   &                     0.2090                 &               0.1641                        & 0.1948                                 & 0.2851                                  & 0.1628                                & 0.1256                               \\ %\hline 
\textbf{RippleNet}~\cite{wang2018ripplenet}                   & 0.2834                                 & 0.4042                                  & 0.2219                                & 0.1780                               & 0.1965                                 & 0.2865                                  & 0.1638                                & 0.1265                               \\ \hline

\textbf{RuleRec$_{two}$} (BPRMF)               & 0.3495*                                 & 0.4768                                  & 0.2813*                                & 0.2201*                               & 0.2050*                                 & 0.2932                                  & 0.1707*                                & 0.1334*                               \\% \hline
\textbf{RuleRec$_{multi}$} (BPRMF)                    & 0.3568*                                 & 0.4829*                                  & 0.2864*                                & 0.2246*                               & 0.2071*                                 & 0.2946*                                 & \textbf{0.1718*}                                & \textbf{0.1341*}                               \\% \hline
\textbf{RuleRec$_{two}$} (NCF)                 & 0.3538*                                 & 0.4876*                                  & \textbf{0.2902*}                       & \textbf{0.2296*}                      & 0.2049*                                 & \textbf{0.2947*}                                   & 0.1681                                & 0.1296                               \\ %\hline
\textbf{RuleRec$_{multi}$} (NCF)           & \textbf{0.3569*}                        & \textbf{0.4894*}                         & \textbf{0.2902*}                       & 0.2290*                               &    \textbf{0.2074*}                                    &     0.2917                                    &     0.1702*                                  &             0.1330                         \\ \hline
\end{tabular}
\end{small}
\label{tab:reco_result}
\vspace{0.0cm}
\end{table*}

\subsection{Experiments and Performance Study}
The experimental results of these algorithms in different domains are summarized in Table~\ref{tab:reco_result}. We repeated each setting for 5 times and conducted the paired two-sample t-test on the 5 times experiment results for significant test. 
As shown in the table, the performance of algorithms in Electronic dataset is obviously worse than in Cellphone dataset. The reason is that the item count of Electronic dataset is about 6 times over the item count of Cellphone dataset (from Table~\ref{tab:reco_data}), which makes the recommendation in Electronic dataset more difficult. 

\smallskip \noindent \textbf{1. The Enhanced Algorithms vs. the Originals.} NCF algorithm performs better than BPRMF algorithm in both datasets, as more complex user and item feature interactions are taken into consideration in NCF. Looking into the results of BPRMF algorithms and NCF algorithms, we find that {RuleRec$_{multi}$} with BPRMF gets 6.5\% to 11.0\% improvements over BPRMF in different evaluation metrics on two domains. The improvements of {RuleRec$_{multi}$} with NCF in Recall@5, Recall@10, NDCG@10, and MRR@10 are between 3.0\% to 6.4\% comparing with NCF  in Cellphone domain, while the improvements of which on Electronic is lower than  in Cellphone domain. 
Though {RuleRec$_{multi}$} with BPRMF is improved more than {RuleRec$_{multi}$} with NCF, {RuleRec$_{multi}$} with NCF still achieves the best performance in Cellphone domain and {RuleRec$_{multi}$} with BPRMF performs the best in Electronic domain.

\smallskip \noindent \textbf{2. Overall Performances.} Besides, we find that any one of the enhanced algorithms outperform all baselines in both Cellphone and Electronic domains in each metric. And most of the improvements are statistically significant, showing that the derived rules from the knowledge graph are really helpful to generate a better ranked item list for the recommendation. 
The multi-task learning algorithms ({RuleRec$_{multi}$} with BPRMF and {RuleRec$_{multi}$} with NCF) show better performances than the two-step learning algorithms ({RuleRec$_{two}$} with BPRMF and {RuleRec$_{two}$} with NCF), indicating that the combination of recommendation loss and rule selection loss in weight training is able to boost the recommendation results. Though the learning-based selection methods perform worse than chi-square in Section 4.2, it does helpful in the multi-task learning.

\smallskip \noindent \textbf{3. The Performances of HERec and RippleNet.} We also note that HERec based algorithms and RippleNet, some state-of-the-art algorithms that uses the knowledge graph for the recommendation, performs worse in these datasets. We think the possible reason is that unlike movie, book, or Yelp datasets which contains many well organized category features (such as director, movie type, actor/actress name in movie dataset) to construct a compact graph, here we link Cellphone and Electronic datasets with a real knowledge graph Freebase. Though Freebase contains more information, but it is not as clean as the on-topic sub graph and makes it harder to mine valuable information, so these algorithms perform worse. More analyses are shown in Section 5.5.1.
% \blue{I think this is a weak point. Review can ask why we don't use other datasets}

To summarize, the derived rules from knowledge graph are valuable for item pair feature vector learning, and the learned vector is able to enhance multiple basic recommendation models (BPRMF and NCF here). Comparing with the two-step combination method, multi-task learning for both recommendation and rule selection contributes more on rule weight learning. Due to the flexible of the  proposed framework, the derived rules are able to combine with other recommendation models to boost performances significantly. 

\subsection{Case Study and Performance Analysis}
% \smallskip
% \noindent
\textbf{1. Performance Comparison in compact heterogeneous graph.} 
Experiments in Section 5.4 are conducted on a large heterogeneous graph extracted from real knowledge graph. In this subsection,  some extra experiments are conducted on a compact heterogeneous graph, which is constructed based on item attributes, in MovieLens-1M dataset\footnote{https://github.com/hwwang55/RippleNet/tree/master/data/movie}.
We adopt the proposed algorithm and HERec algorithm in this dataset following the setting in RippleNet.

The experimental results are shown in Table 6.  Our model performs better than HERec while worse than RippleNet , there are two possible reason: 1) relation type is very limited in this dataset (only 7), so the power of rule selection for the recommendation in RuleRec is limited in this scenario. 2) MovieLens-1M is different from real knowledge graph datasets in Section 5.4 (which is constructed by linking items into Freebase), the connection coverage of it is very perfect and RippleNet benefits a lot from this.
The results indicate that the proposed algorithms is able to achieve noteworthy performance in compact heterogeneous graph.

\begin{table}[]
\caption{\textbf{Performance Comparison on MovieLens dataset.}}
\vspace{-1em}
\begin{tabular}{|c|c|}
\hline
\textbf{Model}         & \textbf{AUC} \\ \hline
% \textbf{CKE}~\cite{zhang2016collaborative}           & 0.796        \\ \hline
% \textbf{PER}~\cite{yu2013recommendation}           & 0.712        \\ \hline
\textbf{Hec$_{sl}$}~\cite{shi2018heterogeneous}        & 0.894        \\ \hline
\textbf{Hec$_{pl}$}~\cite{shi2018heterogeneous}        & 0.895        \\ \hline
\textbf{RippleNet}~\cite{wang2018ripplenet}     & \textbf{0.921}        \\ \hline
\textbf{RuleRec$_{two}$(BPRMF)} & 0.907        \\ \hline
% \textbf{RuleRec$_{two}$(NCF)}   & 0.891        \\ \hline
\end{tabular}
% \vspace{-1em}
\end{table}

\smallskip
\noindent
\textbf{2. Explainability of the learned rules.}
In Section 5.4, the results indicate the derived rules are useful in providing more accurate recommendation results. In this section, we will show the explainability of the derived rules for the recommendation.
Two positive weighted rules on RuleRec$_{multi}$ are shown as the following :
\begin{itemize}
    \item $R_1$ = ``computer.computer.manufacturer"
    \item $R_2$ = ``computer.computer.compatible\_oses" $->$ 
    
    ``computer.os\_compatibility.operating\_system" $->$ 
    
    ``computer.operating\_system.includes\_os\_versions"
\end{itemize}
Where the words with quotation marks are the relation types defined in Freebase (such as ``computer.computer.manufacturer").  These rules are with positive weights in the recommendation module, indicating that if a new item $b$ exists a path between it and item $a$ user bought before, item $b$ is more likely to get higher score. 

First we try to verify if item pairs with these rules affect user's purchase. As to $R_1$, it links a computer product and its manufacturer. If two items $a$ and $b$ have a path in $R_1$, it means that item $b$ is likely to be manufactured by the same as item $a$. For $R_2$, two example entity paths in this rule are: 1) ``Mac Mini" - ``os x yosemite" - ``OS X" - ``IOS" and 2) ``Surface Pro" - ``Windows 10" - ``Windows" - ``Windows Phone". It shows that users are tend to use similar operating systems in both cellphone and computer. As you can see, these rules are consistent with our common sense.

Then, to check whether users agree that the selected rule will be helpful to improve the explainability of the recommendation if the rules are used in real scenarios, the derived rules in Cellphone dataset are labeled by three experts (only agree or disagree, 100 rules from ALB and BT associations). The results show that over 94\% learned rules are accepted by users (87\% rules are accepted by all users). 

Due to the effective rule $i$ in calculating user preference on a specific item  will get higher score ($\mathbf{w_i}^\top{{F}}_{(i, I_u | r_i)}$) for the preference prediction. So for each item in the ranked list, unless it has no path between it and items in user's purchase history, we can generate the most important rule for it by ranking the score of each rule in preference prediction.

\begin{figure*}[htbp]
\centering
\subfigure{
\begin{minipage}[t]{0.24\linewidth}
\centering
\includegraphics[width=\linewidth]{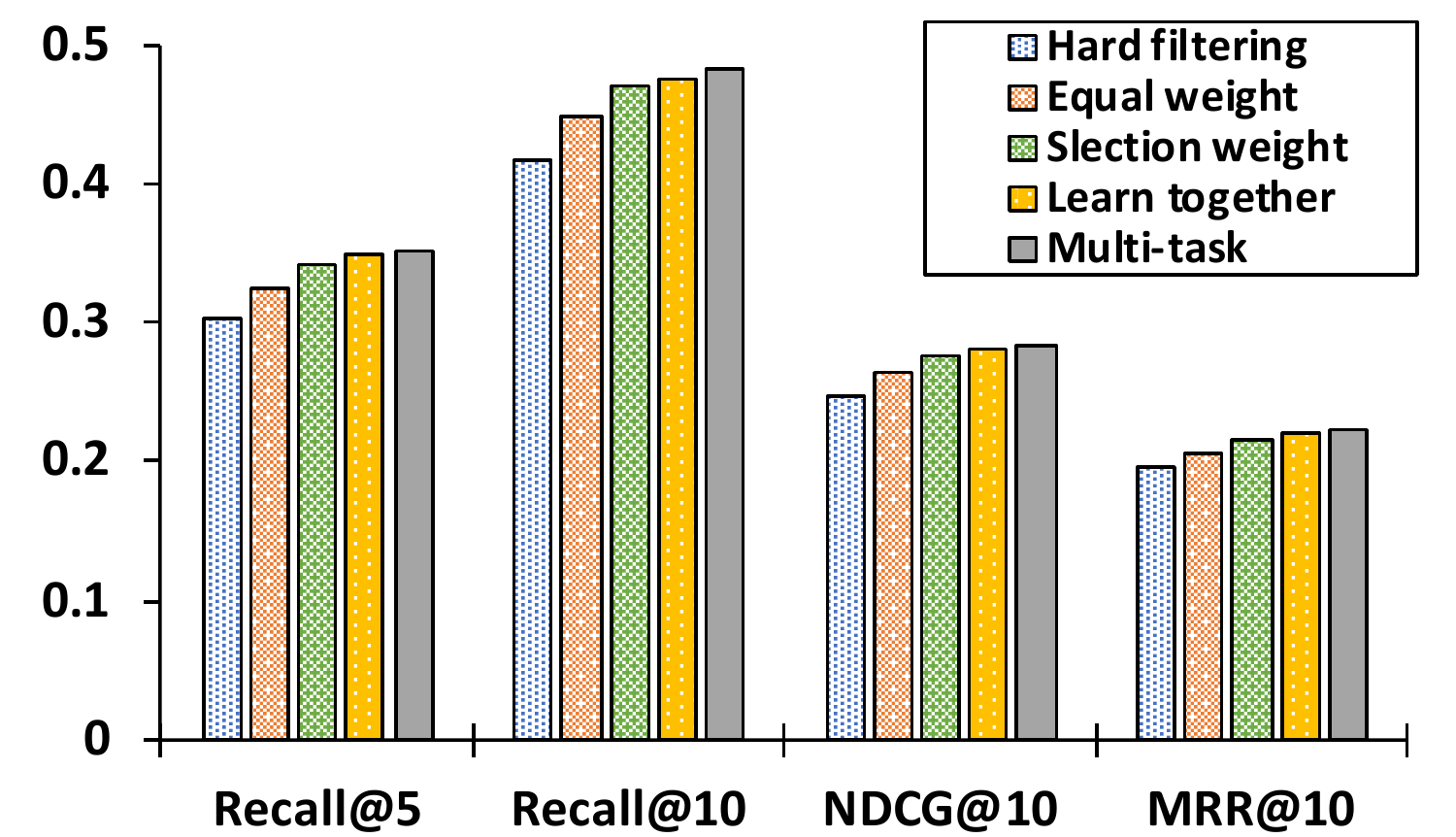}
\text{(a) ALV association}
%\caption{fig1}
\end{minipage}%
}%
\subfigure{
\begin{minipage}[t]{0.24\linewidth}
\centering
\includegraphics[width=\linewidth]{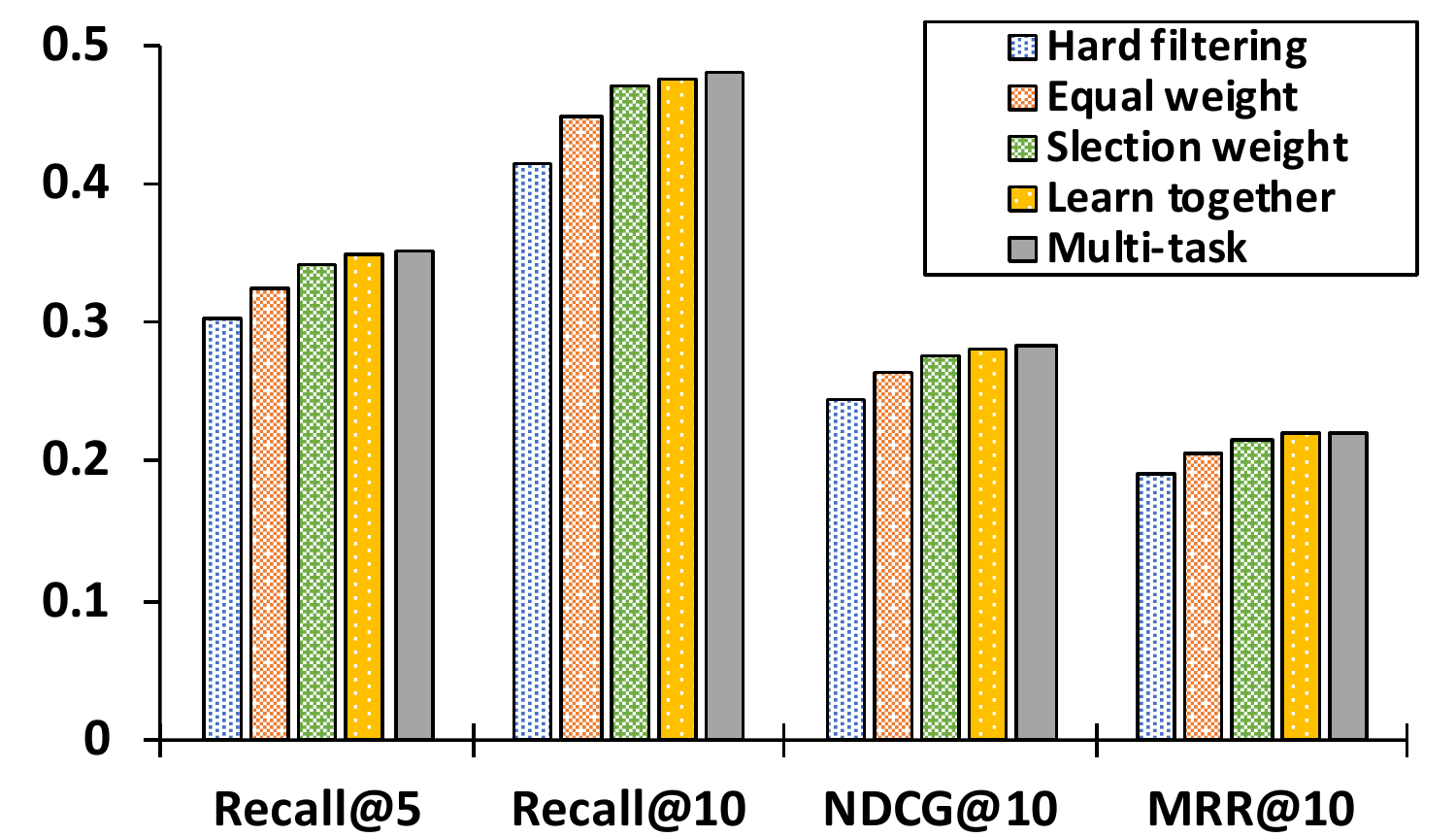}
%\caption{fig2}
\text{(b) BAV association}
\end{minipage}%
}%
\subfigure{
\begin{minipage}[t]{0.24\linewidth}
\centering
\includegraphics[width=\linewidth]{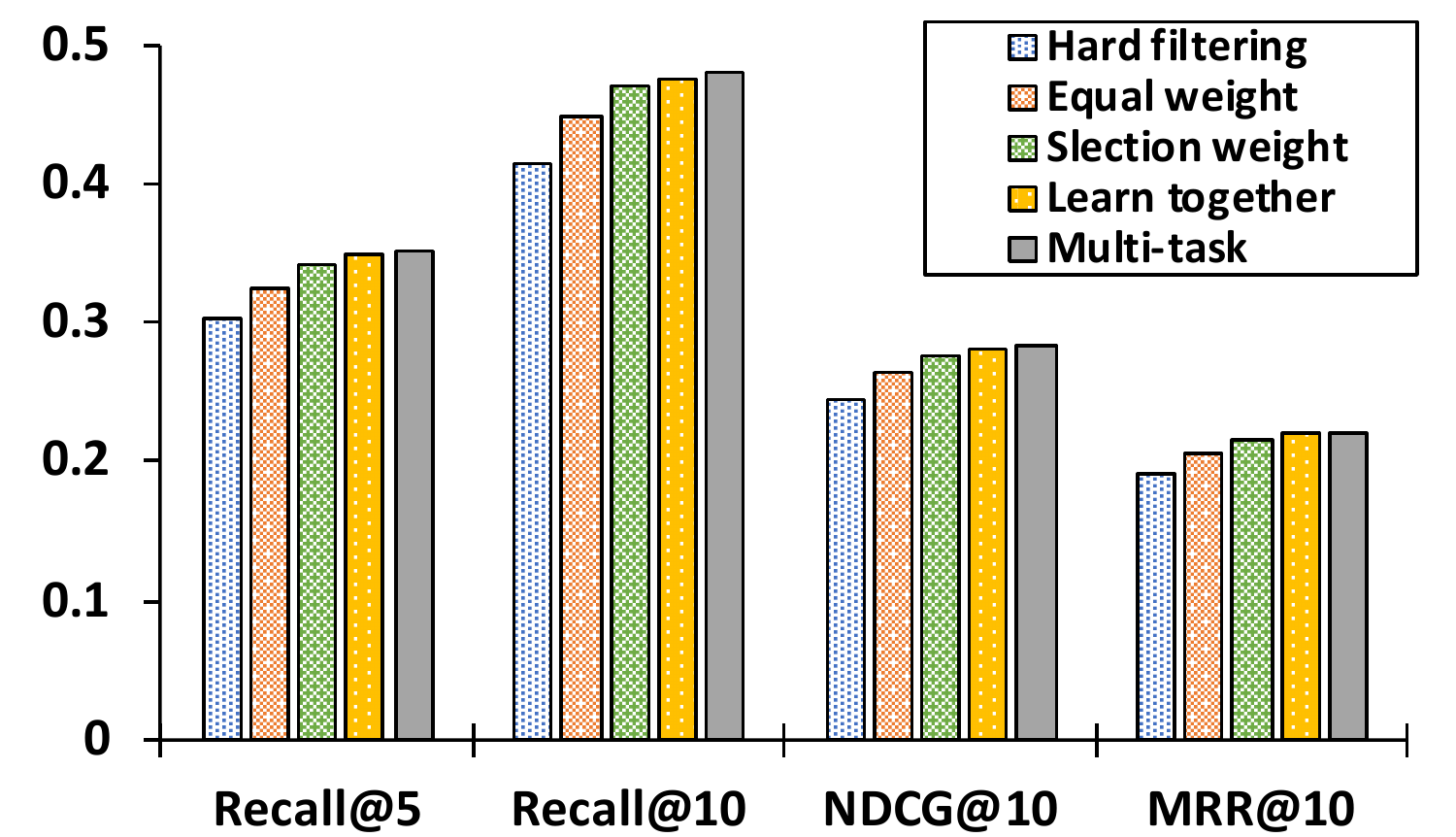}
%\caption{fig2}
\text{(c) ABU association}
\end{minipage}
}%
\subfigure{
\begin{minipage}[t]{0.24\linewidth}
\centering
\includegraphics[width=\linewidth]{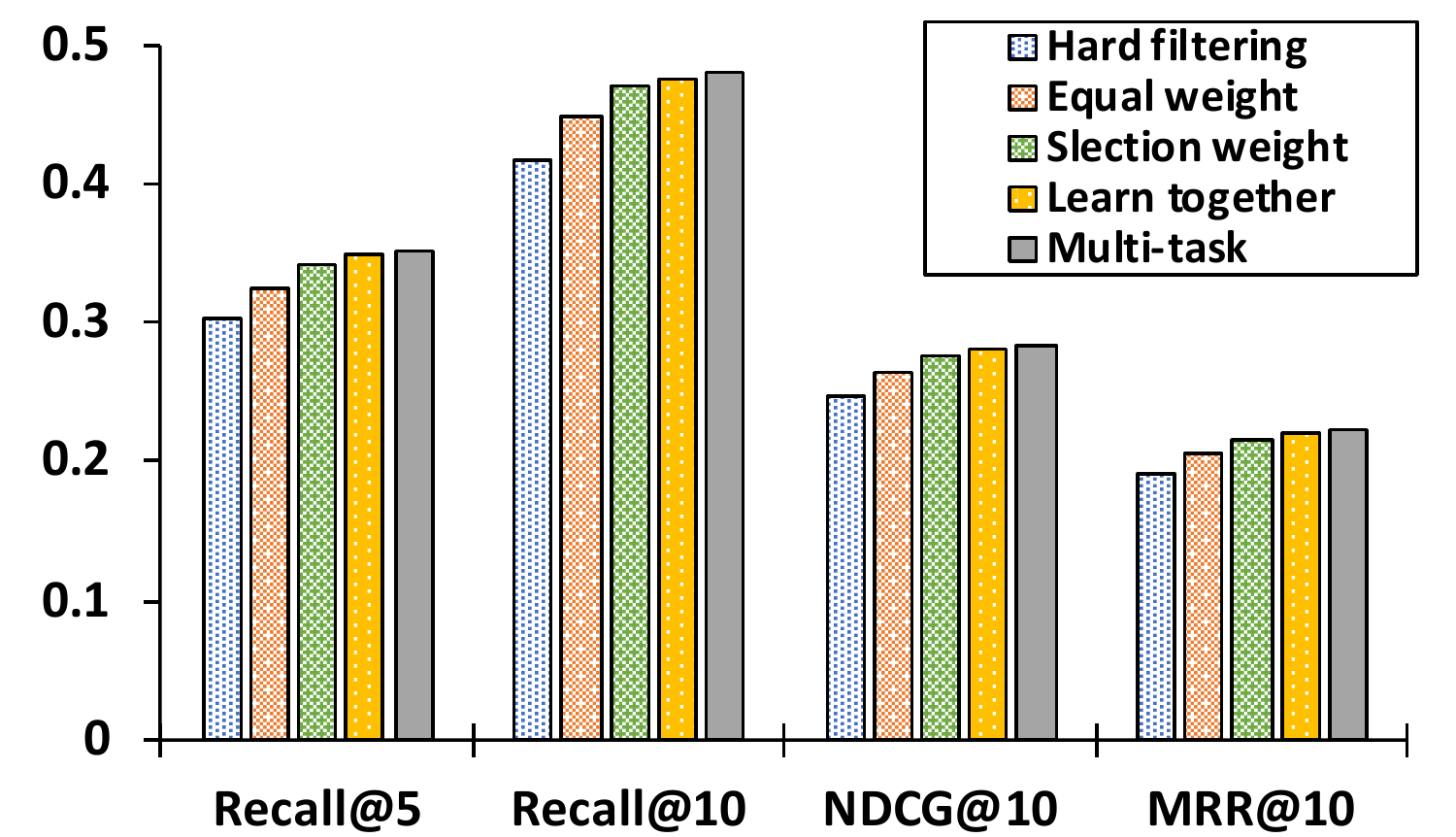}
%\caption{fig2}
\text{(d) BT association}
\end{minipage}
}%
\centering
\vspace{-1em}
\caption{\textbf{Performance comparison under different score function settings in the Cellphone dataset.} We select the top 50 rules in each association by chi-square method.  Among all score function settings, the multi-task learning method performs the best. }
%\blue{fonts are small, and you can use colored bar graph}
\label{fig:func}

\end{figure*}

\smallskip
\noindent
\textbf{3. Study on different model integration strategies.}
\label{sec:comb}
% The implementation of the combined score function \blue{(equation \eqref{})}
The score function (Eq \eqref{score_re}) and %the setting of 
the rule weight vector $\mathbf{w}$ in $f_\mathbf{w}$ affect the performance of the recommendation module. We experiment with several ways to identify the best combination methods.
\begin{itemize}
    \item Hard filtering: Remove candidate items that  have no rule with any item in $I_u$. Formally, $S^{'}_{u,i} =  S_{u,i} \cdot I({{F}}_{(i, I_u | R)})$, where $I({{F}}_{(i, I_u | R)}) = 1$ if $\sum {{F}}_{(i, I_u | R)} >= 1$ otherwise 0. 
    % \item Multiple combine: The score is calculated by multiplying the original score with rule part score. $S^{'}_{u,i} =  S_{u,i} * \mathbf{w}^T * {{F}}_{(i, I_u | R)})$
    \item Equal weight: Each rule gets an equal weight in prediction. $S^{'}_{u,i} = S_{u,i} + \alpha \cdot \mathbf{w}^\top{{F}}_{(i, I_u | R)}$, and $\mathbf{w} = [0.02,...,0.02]$.
    \item Selection weight:  $S^{'}_{u,i} = S_{u,i} + \alpha \cdot \mathbf{w}^\top {{F}}_{(i, I_u | R)}$, and $\mathbf{w} = \mathbf{w}_{rule selection}$ $\mathbf{w}_{rule selection}$ is the rule weight vector trained by Eq~(12) in rule selection step.
    \item Learn together: The rule weight vector is trained with the original recommendation model. $f_\mathbf{w}(a, b) = a + \mathbf{w}^\top b$.
    \item Multi-task: The rule weight vector $\mathbf{w}$ is shared by recommendation learning and rule selection part, and the score prediction function is $f_\mathbf{w}(a, b) = a + \mathbf{w}^\top b$.
\end{itemize}
% $S^{'}_{u,i} =  S_{u,i} \cdot I({{F}}_{(i, I_u | R)})$, where $I({{F}}_{(i, I_u | R)}) = 1$ if $\sum {{F}}_{(i, I_u | R)} >= 1$ otherwise 0.

% $S^{'}_{u,i} = S_{u,i} + \alpha \cdot {w}^\top{{F}}_{(i, I_u | R)}$, and ${w} = [0.02,...,0.02]$

% $S^{'}_{u,i} = S_{u,i} + \alpha \cdot {w}^\top {{F}}_{(i, I_u | R)}$, and ${w} = {w}_{rule selection}$

% The rule weight vector is trained with the original recommendation model. $f_{w}(a, b) = a + {w}^\top b$

% The rule weight vector ${w}$ is shared by recommendation learning and rule selection part

The results of applying multiple score functions  on RuleRec(BPRMF) model in Cellphone dataset are shown in Fig.~\ref{fig:func}. The performances of the score functions are Hard filtering $<$ Equal weight $<$ Selection weight $<$ Learn together $<$ Multi-task with all metrics in all associations. 
First, we can see that the multi-learning method achieves the best performance on all metrics, and the improvements are significant, which indicates multi-task learning is very helpful in rule weight learning for better recommendation results. 
Second, since the hard filtering method is likely to ignore both negative items and positive items (from Table \ref{fig:rule_sel}, we can see that sometimes there is no rule between the positive item and item purchase history $I_u$). 
Third, though selection weight contributes on the recommendation (better that equal weight), it is still worse than Learn together model.

\begin{figure*}[htbp]
\vspace{0.0cm}
\centering
\subfigure{
\begin{minipage}[t]{0.24\linewidth}
\centering
\includegraphics[width=\linewidth]{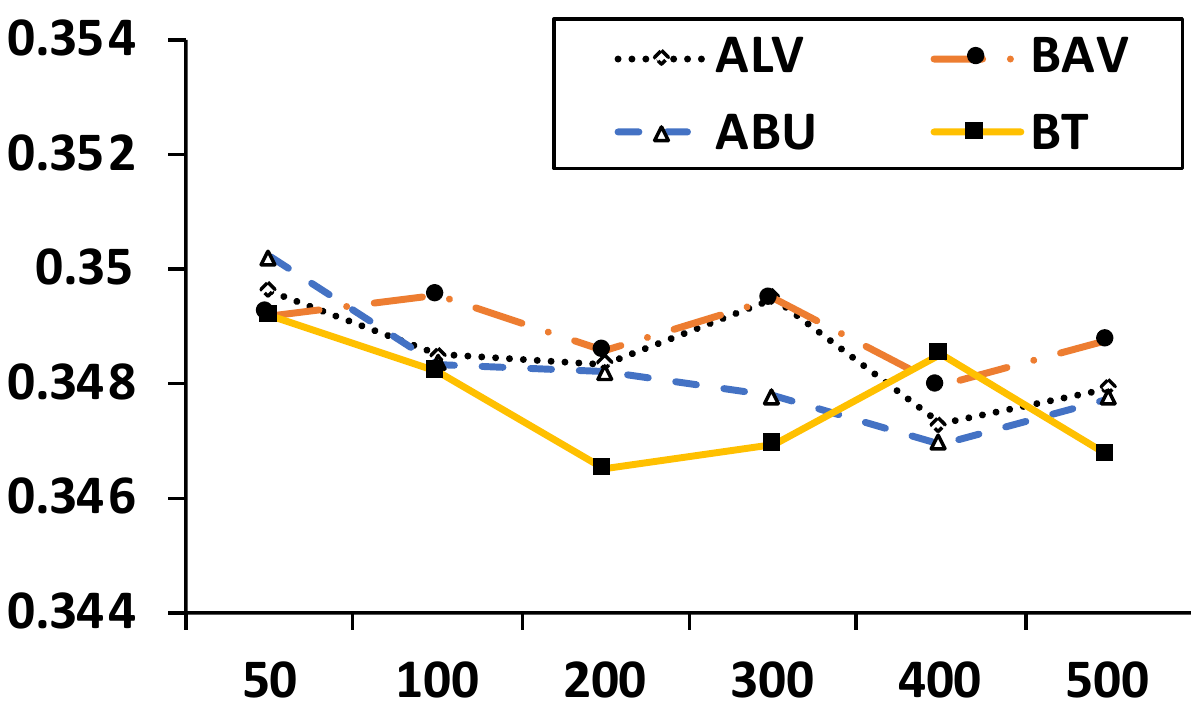}
%\caption{fig1}
\text{(a) Recall@5}
\end{minipage}%
}%
\subfigure{
\begin{minipage}[t]{0.24\linewidth}
\centering
\includegraphics[width=\linewidth]{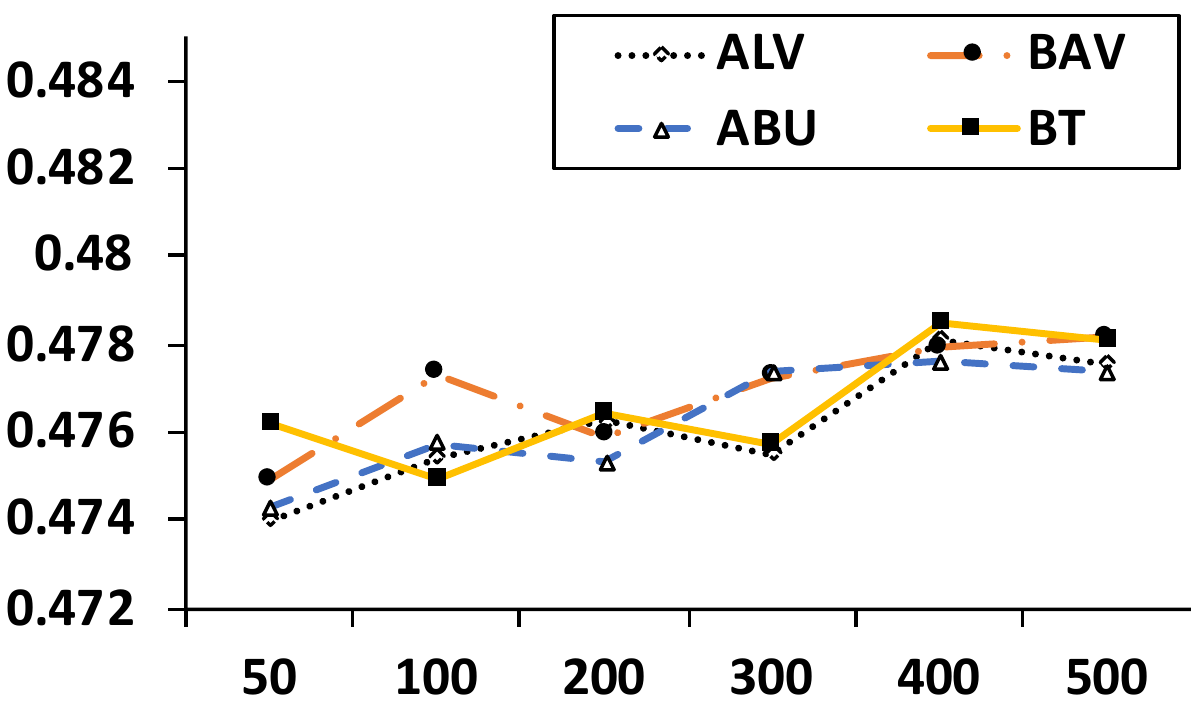}
%\caption{fig2}
\text{(b) Recall@10}
\end{minipage}%
}%
\subfigure{
\begin{minipage}[t]{0.24\linewidth}
\centering
\includegraphics[width=\linewidth]{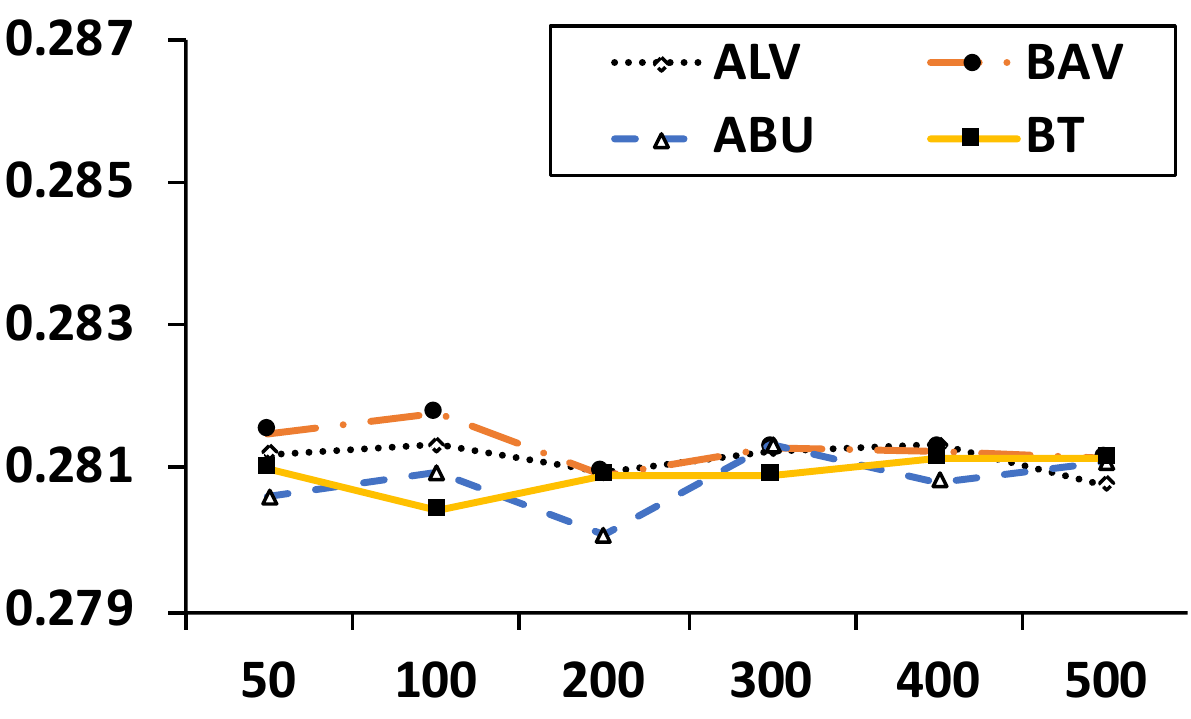}
%\caption{fig2}
\text{(c) NDCG@10}
\end{minipage}
}%
\subfigure{
\begin{minipage}[t]{0.24\linewidth}
\centering
\includegraphics[width=\linewidth]{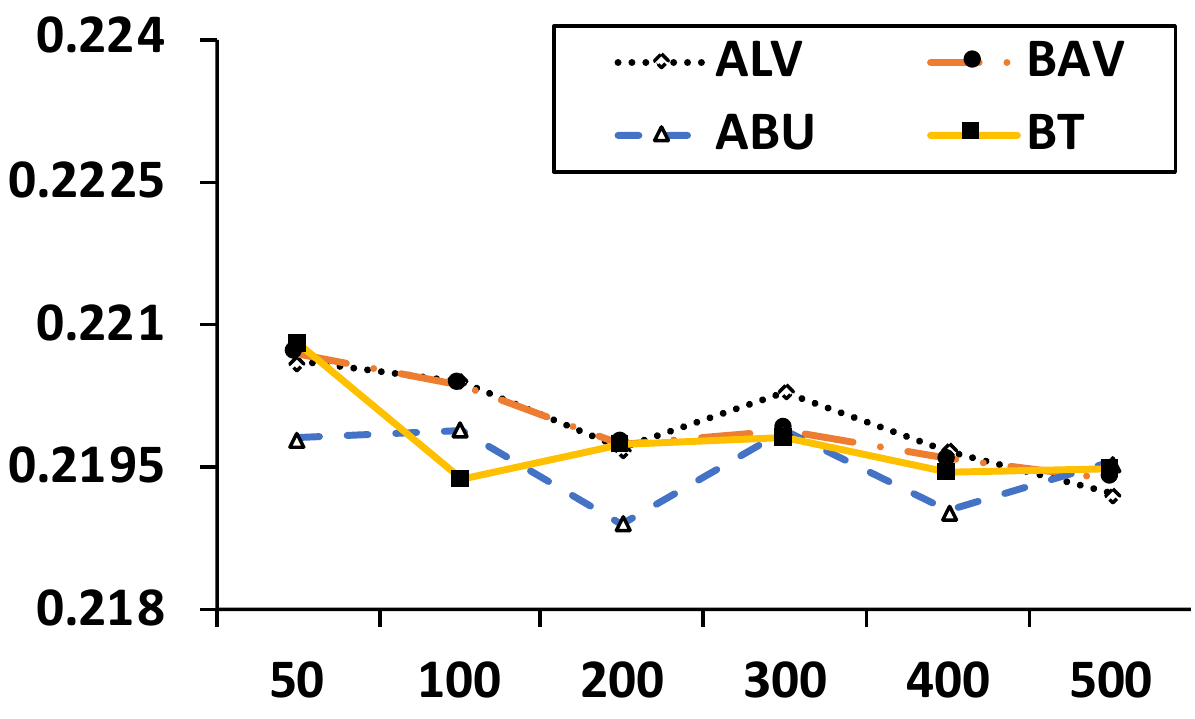}
%\caption{fig2}
\text{(d) MRR@10}
\end{minipage}
}%
\centering
\label{fig:rule_count:two}
\vspace{-0.2cm}
\caption{The performance of using different number of rules in RuleRec$_{two}$ with BPRMF.}
\vspace{-0.3cm}
\end{figure*}

\begin{figure*}[htbp]
\centering
\subfigure{
\begin{minipage}[t]{0.24\linewidth}
\centering
\includegraphics[width=\linewidth]{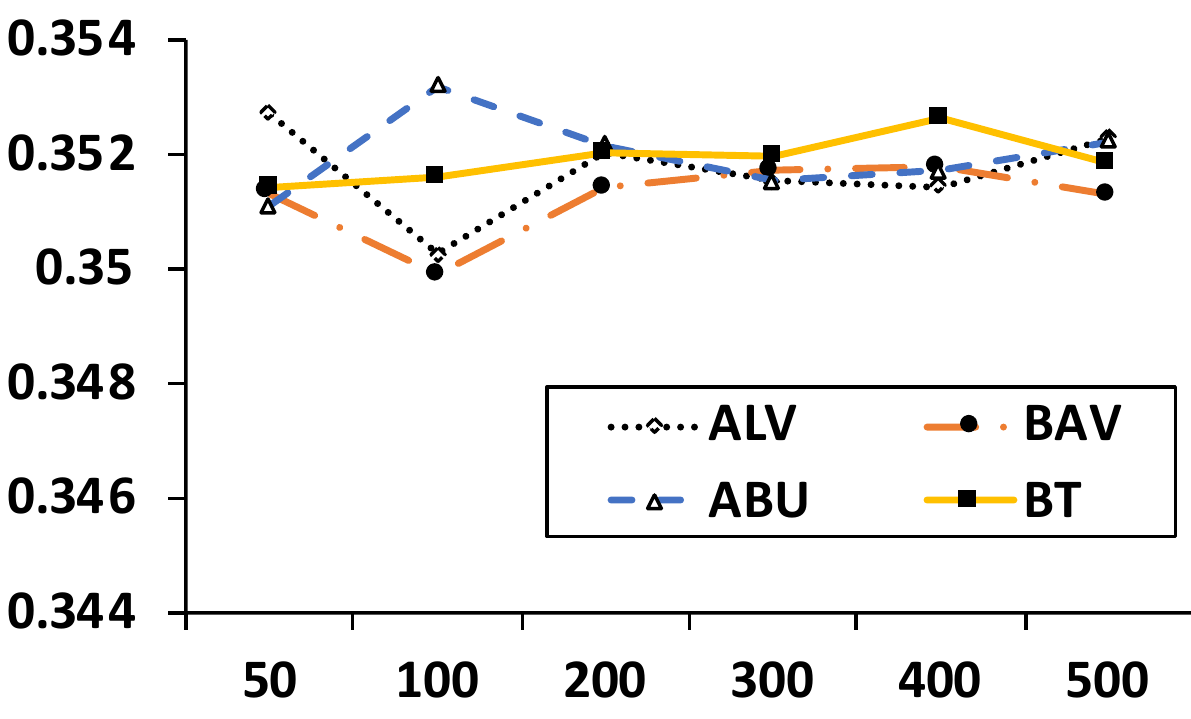}
%\caption{fig1}
\text{(a) Recall@5}
\end{minipage}%
}%
\subfigure{
\begin{minipage}[t]{0.24\linewidth}
\centering
\includegraphics[width=\linewidth]{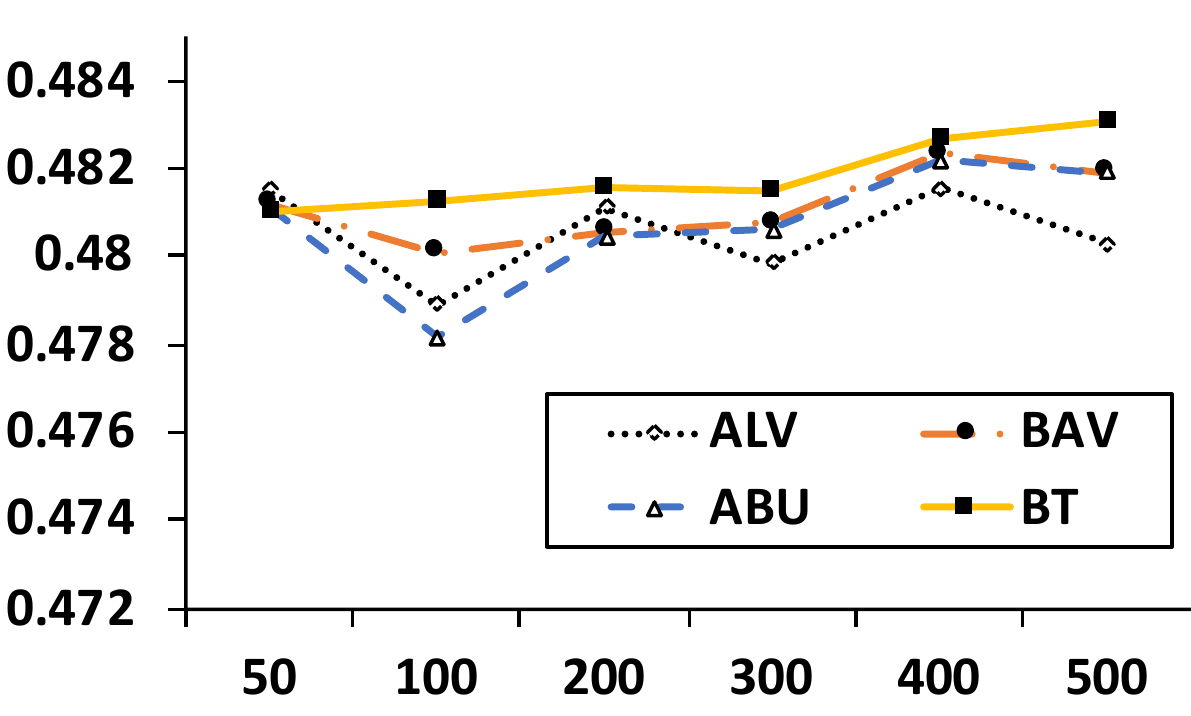}
%\caption{fig2}
\text{(b) Recall@10}
\end{minipage}%
}%
\subfigure{
\begin{minipage}[t]{0.24\linewidth}
\centering
\includegraphics[width=\linewidth]{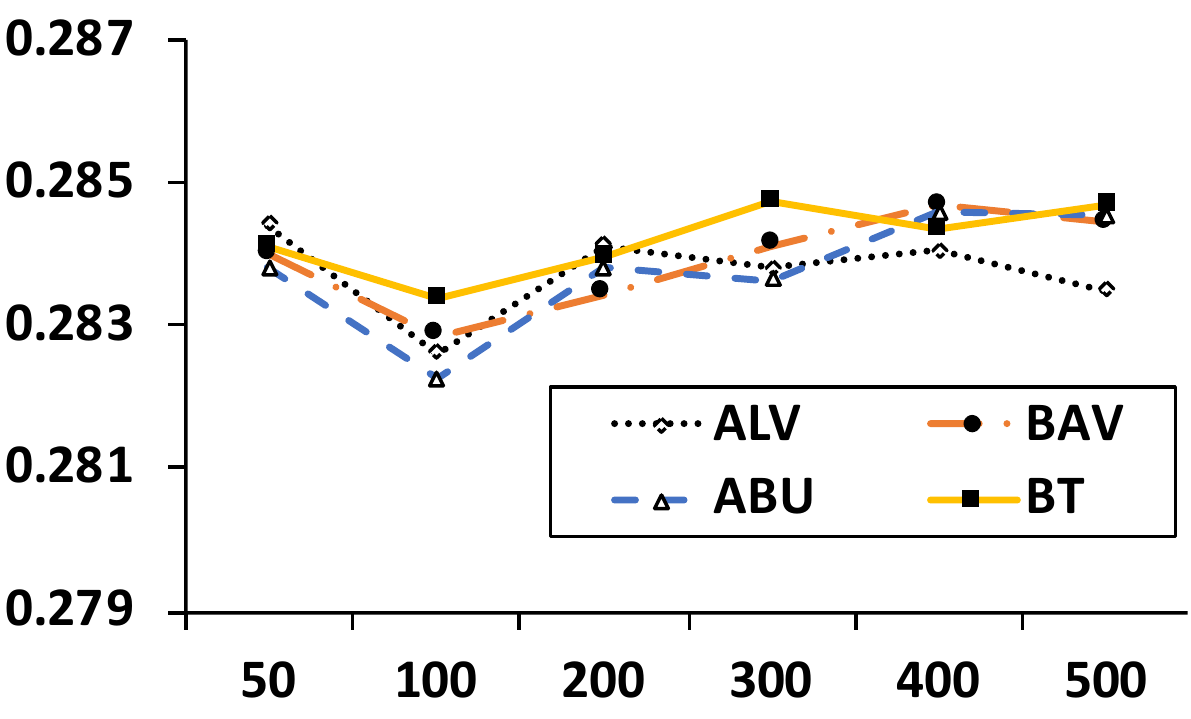}
%\caption{fig2}
\text{(c) NDCG@10}
\end{minipage}
}%
\subfigure{
\begin{minipage}[t]{0.24\linewidth}
\centering
\includegraphics[width=\linewidth]{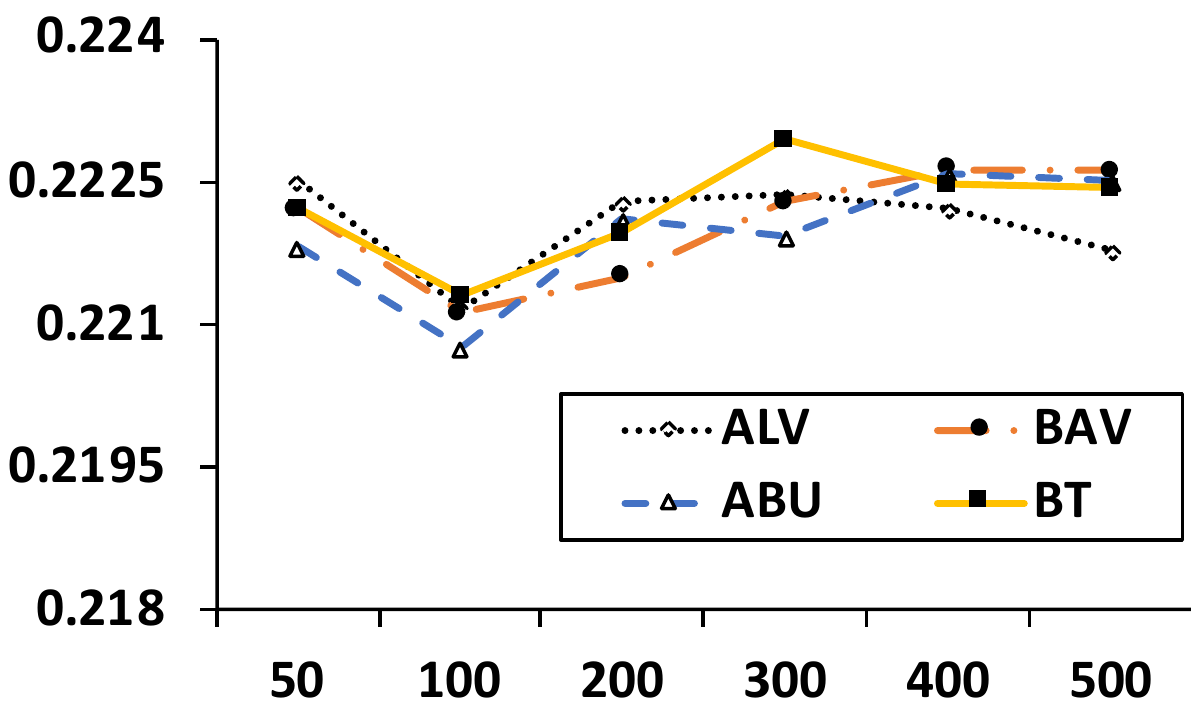}
%\caption{fig2}
\text{(d) MRR@10}
\end{minipage}
}%
\centering
\label{fig:rule_count:multi}
\vspace{-0.2cm}
\caption{The performance of using different number of rules in RuleRec$_{multi}$ with BPRMF.}
% \blue{line width is narrow and the point is too small. Fonts are also small.}
\vspace{-0.2cm}
\end{figure*}

\smallskip
\noindent
\textbf{4. Study on single association vs. all associations}
\label{sec:assoc}

In this subsection, we compare the performance of RuleRec$_{multi}$ with BPRMF with only one type of association and all associations, the results are summarized in Table~\ref{tab:associa}.

\begin{table}[]
\caption{\textbf{The results of using a single association vs. all associations.} BPRMF is used as a recommendation model. Our proposed model shows the best performances when using all kinds of associations.}
\begin{tabular}{l|cccc}
\hline
\textbf{Type} & \multicolumn{1}{c}{\textbf{Recall@5}} & \multicolumn{1}{c}{\textbf{Recall@10}} & \multicolumn{1}{c}{\textbf{NDCG@10}} & \multicolumn{1}{c}{\textbf{MRR@10}} \\ \hline
\textbf{None}         & 0.3238                                 & 0.4491                                  & 0.2639                                & 0.2058                               \\
\textbf{ALV}         & 0.3527                                 & 0.4815                                  & 0.2844                                & 0.2225                               \\ 
\textbf{BAV}         & 0.3513                                 & 0.4812                                  & 0.2840                                & 0.2222                               \\ 
\textbf{ABU}         & 0.3511                                 & 0.4811                                  & 0.2838                                & 0.2218                               \\ 
\textbf{BT}          & 0.3514                                 & 0.4810                                  & 0.2841                                & 0.2222                               \\ 
\textbf{ALL}         & \textbf{0.3568}                                 & \textbf{0.4829}                                  & \textbf{0.2864}                                & \textbf{0.2246}                               \\ \hline
\end{tabular}
% \vspace{-1em}
\label{tab:associa}
\vspace{-1em}
\end{table}

First, we can see that with the rules derived by any one of the four associations, {RuleRec$_{multi}$} with BPRMF outperforms BPRMF algorithm significantly. The performances of using different associations are similar, but all of them are valuable for mining the item relationships to boost the recommendation results. 
Second, RuleRec$_{multi}$ with BPRMF derived by all kinds of associations outperforms RuleRec$_{multi}$ with a single association, indicating that the combination  contributes for the recommendation models.

% \subsection{Rule Selection Loss}
% may remove it 

\smallskip
\noindent
\textbf{5. Recommendation with different rule counts.}
\label{sec:rulecounts}
In Section~\ref{sec:rule_module}, rule selection is introduced as an important part in rule learning. Does rule selection is really necessary? We conduct further experiments on each association with different of derived of rules (selected with chi-square method, 50, 100, 200, 300, 400, and 500 respectively) with RuleRec$_{two}$ and RuleRec$_{multi}$ with BPRMF.
% are adopted here. 

From Fig.~7, the performance of RuleRec$_{two}$ with BPRMF decreases in recall@5, MRR@10 as the rule number increases. At the same time, NDCG@10 keeps stable and Recall@10 increases. The overall performances is not getting better when more rules are applied in the recommendation learning. The possible reason is that with the grows of rule number, lots of ''bad" rules are included and the two-step model RuleRec$_{two}$ with BPRMF shows worse ability in dealing with them properly. However, due to the rule selection is taken into consideration in the multi-task learning based algorithms, we find that the performance of RuleRec$_{multi}$ with BPRMF algorithm shows better performances
% keeps good and even better than before 
as the rule number increases (Fig.~8).
Furthermore, the performances of RuleRec$_{multi}$ with BPRMF is significantly better than those of RuleRec$_{two}$ with BPRMF (paired two-sample t-test on the experimental results with different count of rules. $p < 0.01$). 
The results show that the multi-task learning based algorithms are able to tackle with large number of rules even though there are useless rules. 
  
\label{experiments}

%\vspace{-0.1cm}
\section{Related Work}
\textbf{Combine Side-information for the Recommendation.}
Matrix factorization based algorithms \cite{rennie2005fast, rendle2009bpr} are widely used to tackle recommendation problems. Recently, recommendation algorithms achieve remarkable improvements during these years with help of deep learning models \cite{he2017neural, shen2018interactive, zhao2018deep, he2017nfm, zhang2018discrete} and the successful introducing of side-information \cite{ma2018your, wang2018ripplenet, mcinerney2018explore, hu2018interpretable, nandanwar2018fusing}. In this study, we focus on the introducing of side-information in the knowledge graph for the recommendation, and there already two types of studies using the knowledge graph in the recommendation: path-based and embedding learning based. 

Path-based methods adopt random walk on predefined meta-paths between user and items in the knowledge graph to calculate user's preference on an item. Yu et al. first propose to use meta-paths to utilize user-item preferences and then expand matrix factorization for the recommendation \cite{yu2013recommendation}. Shi et al. use weighted paths for explicit recommendation \cite{shi2015semantic}. Zhao et al. design a factorization machine with the latent features from different meta-paths \cite{zhao2017meta}. Catherine et al. design a first-order probabilistic logical reasoning system, named ProPPR, to integrate different meta-paths in a knowledge graph \cite{catherine2017explainable, catherine2016personalized}.  All these methods achieve improvements in the recommendation, while the weakness of them is that they ignore the type of item associations.

Embedding learning based methods conduct user/item representation learning based on the Knowledge graph structure firstly. The learned embedding \cite{grover2016node2vec} is applied in Zhang's study to get item embedding for the recommendation \cite{palumbo2017entity2rec}. Zhang et al. use TransR \cite{lin2015learning} to learn the structural vectors of items, and these vectors are part of the final item latent vector for preference prediction \cite{zhang2016collaborative}. Besides, some previous studies propose new algorithms in which Meta-path guided random walks are used in a heterogeneous network for user and item embedding learning and achieve outperform results \cite{shi2018heterogeneous,zhao2017meta, wang2018path, wang2018ripplenet} in different ways.
However, the embedding learning based methods give up the explainable strength of the knowledge graph, which is very valuable for the recommendation.

\smallskip
\noindent
\textbf{Rule Learning in the Knowledge Graph.}
Item-item relationships are considering as useful features for providing better recommendation results. Julian et al. firstly propose to a topic model based method to predict relationships (substitute or complementary) between products from reviews \cite{mcauley2015inferring}. In their study, the ground truth is calculated in a data-driven way in Amazon. Then, more algorithms attempt to improve the  prediction results with better algorithms. Word dependency paths are taken into consideration in Hu et al.'s work \cite{xu2016cer} and Wang et al. adopt a embedding based method to enhance the performance of relationship prediction \cite{wang2018path}.

However, these methods are suffering from cold items. 
% And these associations (e.g. ``also view'') cannot directly infer if two items are substitute items or complementary items. 
% For example, you may browse different laptops when you want to buy one, and you may also browse laptops and other items (such as a keyboard) but only buy laptop. However, in such condition, the keyboard will be seen as a substitute item of the laptop. Obviously it is not the truth. 
So we proposed to not predict item associations directly, but mine meaningful rules in  the knowledge graph  with the ground truth item pairs. The rules will be applied to  generate feature vectors for different item pairs without user reviews. % and the used in the recommendation part.

Knowledge graph is a multi-relational graph that composed of entities as nodes and relations as different types of edges \cite{wang2014knowledge}. In the past years, knowledge graphs have been used as important resources for many tasks  \cite{xiong2017explicit, dubey2018earl}. One of the main usage of knowledge graph is reasoning and entity relationship prediction. Lots of research are focus on reasoning, such as  \cite{xiong2017deeppath,yang2017differentiable}. While these studies focus on link prediction but not rule inducing, which is not proper for our study.

On the other line of research, several work attempt to learn the useful rules but not the prediction results from the knowledge graph with the ground truth entity pairs. Random walk based algorithms are proposed in Lao et al.'s studies \cite{lao2010relational, lao2011random} and others'  \cite{wang2015knowledge, guo2016jointly}. These methods are able to show why the entity pair has a certain relationship according to the derived rules, which makes the results more explainable. In our study, we adopt a similar algorithm as them in rule learning module.

\label{related}

\vspace{-0.1cm}
\section{Conclusions and Future Work}
In this paper, we propose a novel and effective joint optimization framework for inducing rules from a knowledge graph with items and recommendation based on the induced rules. 

Our framework consists of two modules: rule learning module and recommendation module. The rule learning module is able to derive useful rules in a knowledge graph with  different type of item associations, and the recommendation module introduces the rules to the recommendation models for better performance. Furthermore, there are two ways to implement this framework: two-step and jointly learning. 

Freebase, a large-scale knowledge graph, is used  for rule learning in this study. The framework is flexible to boost different recommendation algorithms. 
We modify two recommendation algorithms, a classical matrix factorization algorithm (BPRMF) and a state-of-the-art neural network based recommendation algorithm (NCF), to combine with our framework. 
The proposed four rule enhanced recommendation algorithms achieve remarkable results in multiple domains and outperform all baseline models, indicating the effectiveness of our framework. Besides, the derived rules also show the ability in explaining why we recommend this item for the user, boosting the explainability of the recommendation models at the same time. Further analysis shows that our multi-task learning based combination methods (RuleRec$_multi$ with BPRMF and RuleRec$_two$ with NCF) outperform the two-step method with different number of rules. And the combination of rules derived by different associations contributes to better recommendation results.

In future, we plan to investigate how to design a embedding learning based combination algorithm which keeps the recommendation results explainable with the knowledge graph. 

\section*{Acknowledgements}
This work is supported by Natural Science Foundation of China (Grant No. 61672311, 61532011) and The National Key Research and Development Program of China (2018YFC0831900). Dr. Xiang Ren has been supported in part by NSF SMA 18-29268, Amazon Faculty Award, and JP Morgan AI Research Award.
\label{conclusion}


\end{document}